\documentclass{wsjai}

\usepackage{txfonts}
\usepackage{graphicx}
\usepackage{url}
\usepackage{hyperref}
\usepackage{stmaryrd}
\usepackage{color}
\definecolor{darkblue}{rgb}{0.0, 0.0, 0.5}
\definecolor{darkred}{rgb}{0.8, 0.0, 0.0}

\hypersetup{colorlinks=true,citecolor=darkblue,linkcolor=darkblue}

\usepackage{setspace}
\def\aj{AJ}%
\def\apj{ApJ}%
\def\apjl{ApJ}%
\def\apjs{ApJS}%
\def\aap{A\&A}%
\def\aaps{A\&AS}%
\def\mnras{MNRAS}%
\def\pasp{PASP}%
\def\s4g{S$^4$G}

\def\nborg{93106}

\def\borgarea{209.9}

\def\Mdwarfs{274}
\def\Ldwarfs{30}

\def\Nfields{71}
\def\f125w{{\em F125W}}
\def\f160w{{\em F160W}}
\def\f098m{{\em F098M}}
\def\f606w{{\em F606W}}
\def\f600lp{{\em F600LP}}

\def\nstars{4448}
\def\Mdwarfs{1410}
\def\Ldwarfs{64}

\def\nzeightgal{50} 

\newcommand{\farcm}{\mbox{\ensuremath{.\mkern-4mu^\prime}}}%    % fractional arcminute symbol: 0.'0
\newcommand{\farcs}{\mbox{\ensuremath{.\!\!^{\prime\prime}}}}%  % fractional arcsecond symbol: 0.''0

\begin{document}

\catchline{}{}{}{}{} % Publisher's Area please ignore

\markboth{Holwerda et al.}{NIRspec Target Acquisition from WFC3 pre-images}

\title{All NIRspec needs is HST/WFC3 pre-imaging?\\
The use of Milky Way Stars in WFC3 Imaging to Register NIRspec MSA Observations}

\author{B.W. Holwerda$^{1}$, R. J. Bouwens$^{1}$,  
M. Trenti$^{2}$, 
and M.A. Kenworthy$^{1}$}

\address{
$^1$ University of Leiden, Sterrenwacht Leiden, Niels Bohrweg 2, NL-2333 CA Leiden, The Netherlands\\
$^2$School of Physics, University of Melbourne, VIC 3010, Australia}

\maketitle

\corres{$^\dagger$Corresponding author.}

\begin{history}
\received{(to be inserted by publisher)};
\revised{(to be inserted by publisher)};
\accepted{(to be inserted by publisher)};
\end{history}

\begin{abstract}
The James Webb Space Telescope (JWST) will be an exquisite new near-infrared observatory with imaging and multi-object spectroscopy through ESA's NIRspec instrument with its unique Micro-Shutter Array (MSA), allowing for slits to be positioned on astronomical targets by opening specific $0\farcs2$-wide micro shutter doors. 

To ensure proper target acquisition, the on-sky position of the MSA needs to be verified before spectroscopic observations start. An onboard centroiding program registers the position of pre-identified guide stars in a Target Acquisition (TA) image, a short 
pre-spectroscopy exposure without dispersion (image mode) through the MSA with all shutters open.

The outstanding issue is the availability of Galactic stars in the right luminosity range for TA relative to typical high redshift targets. 
We explore this here using the stars and $z\sim8$ candidate galaxies identified in the source extractor catalogs of Brightest of Reionizing Galaxies survey (BoRG[z8]), a pure-parallel program with Hubble Space Telescope Wide-Field Camera 3.

We find that
(a) a single WFC3 field contains enough Galactic stars to satisfy the NIRspec astrometry requirement (20 milli-arcseconds), provided its and the NIRspec TA's are $m_{lim}>24.5$ AB in WFC3  {\em F125W}, 
(b) a single WFC3 image can therefore serve as the pre-image if need be,
(c) a WFC3 mosaic and accompanying TA image satisfy the astrometry requirement at $\sim23$ AB mag in WFC3  {\em F125W}.
(d) no specific Galactic latitude requires deeper TA imaging due to a lack of Galactic stars, and
(e) a depth of $\sim24$ AB mag in WFC3 {\em F125W} is needed if a guide star in the same MSA quadrant as a target is required.

We take the example of a BoRG identified $z\sim8$ candidate galaxy and require a Galactic star within 20" of it. In this case, a depth of 25.5 AB in {\em F125W} is required (with $\sim$97\% confidence).

\end{abstract}

%% Keywords should appear after the \end{abstract} command. The uncommented
%% example has been keyed in ApJ style. See the instructions to authors
%% for the journal to which you are submitting your paper to determine
%% what keyword punctuation is appropriate.


\section{\label{s:intro}Introduction}

The James Webb Space Telescope (JWST) will be a game-changing space observatory for high-redshift ($z=2-20$) galaxies. However, it is designed with a limited life-span and no capacity to expand beyond the consumables (propulsion) limit of $\sim$10 years. Therefore a substantial effort is being made to ensure that its observing time is spent as efficiently as possible. JWST's instrument package features two cameras and a spectrograph. 

% NIRspec description
The Near-Infrared Spectrograph (NIRSpec) is the first near infrared multi-object spectrograph in space, capable of observing over a hundred sources in its field-of-view (FOV) of $3\farcm5 \times 3\farcm5$. Targets in the FOV are normally selected by opening groupings of three shutters in the micro-shutter array (MSA) to form an aperture ``slitlet". The microshutters are arranged in a grid that contains more than 62000 shutters with each shutter measuring 100 $\mu$m $\times$ 200 $\mu$m or 0\farcs2 wide and 0\farcs46 on the sky. Multi-object spectroscopy is possible in three spectral resolutions ( R $\sim$ 100, 1000, 2700) over a 0.7-5$\mu$m wavelength range. Additional modes are Fixed-slit and Integral-field spectroscopy but we do not consider these here. The NIRspec instrument needs accurate positions of the target(s) because the width of the slitlets on the sky is similar to the size of the typical targets. 

Thus, one of the efficiency issues is how to ensure the faint, high-redshift target is placed exactly in a ``slitlet" of the NIRspec/MSA during MOS mode without spending much observatory time on pre-imaging, or target acquisition overhead. For the NIRspec/MSA Target Acquisition (TA hereafter), it will be necessary to take a TA image with NIRspec with all the MSA doors open to verify the initial pointing by the telescope using Galactic stars.

It is essential that the MSA slitlets are positioned squarely on the targets to ensure the highest signal-to-noise observed spectrum. 
% For the highest fidelity work, a 
A 20 milli-arcsecond accuracy (10\% of a slitlet) is desirable, which should require some recent and high-quality pre-imaging. 
In this baseline scenario, the typical observing sequence would involve:  
(1) NIRcam pre-imaging in a wide filter (F140X or F110W) in the 1-5 micron wavelength range, 
(2) wait for an interval of several months to process the pre-imaging data and identify targets and reference sources, 
(3) re-acquire the field for JWST/NIRspec observations,  
(4) take a NIRspec pre-spectroscopy image with all MSA slits open,
(5) identify the reference sources onboard and solve for the astrometric solution of the field, and
(6) open the correct slitlets for the spectroscopic observations of the targets.
The pre-imaging step (steps 1 \& 2) is expected to add up to 100 \% to the total cost of observing and necessitate two separate field acquisitions, adding considerable overhead. We refer the reader to the \citet{Beck15,Karakla15} and \citet{Ubeda15} presentations at the ``Exploring the universe with JWST", 49th ESLAB symposium, held at ESTEC in 2015 (\url{http://www.cosmos.esa.int/web/jwst/conferences/jwst2015}) and the JWST technical documents available at the STSCI website: \citet{Regan05} and \citet{Beck09}.

The possibility of pre-imaging with HST would enhance the scientific efficiency of the JWST telescope. 
% We ask whether pre-imaging with HST could streamline the aforementioned observing process with JWST. 
Of particular relevance would be pre-imaging with WFC3/IR Ðas it already provides wavelength overlap with the sensitivity range of NIRSPEC. Such imaging would help by eliminating the time expenditure in steps 1 \& 2 of the above TA process and reduce the number of field acquisitions to one. 
Such a scenario --the use of just WFC3/IR pre-imaging-- has already been considered for HUDF or the CANDELS fields and  
a mask tool to plan NIRspec/MSA observations based  on catalogues from
WFC3 or ACS imaging is being developed. For a mosaic IFU observations one could also consider ground based imaging - but in general,
this will prove to be difficult for NIRspec/MSA work, considering the strict astrometric accuracy required.
%We ask whether pre-imaging with HST could streamline the aforementioned observing process with JWST. Of particular relevance would be pre-imaging with WFC3/IR --as it already provides wavelength overlap with the sensitivity range of NIRSPEC.   Such imaging would help by reducing the time expenditure in steps 1 \& 2 of the above process and reduce the number of field acquisitions to one.

% Holwerda+ 2016, in prep

In considering the above question, the most valuable sources to achieve optimal pointing accuracy are compact sources, specifically stars.
Given the sensitivity range of JWST, the most common stellar sources for this purpose will be red and brown dwarfs. Other types of stars are likely too sparse and/or easily saturated in the target acquisition image (step 4 \& 5) for an astrometric solution. While we concede that it
may be possible that other compact sources, e.g, elliptical galaxies also help in this regard, we will focus our attention on the stellar sources due to their extreme compactness and simplicity of their centeroiding. 

% However, due to the sensitivity of the instrument to light in the mid-infrared and their relative number, the  stars in a given pre-image are likely to be faint dwarf stars --Brown Dwarfs-- for which there will be little information available from all-sky surveys. Other stars are likely too few and easily saturated in the pre-image for an astrometric solution.

% The questions we will try to address are:
This raises the following questions:
\begin{enumerate}
\item Are there enough reference stars within a single WFC3 or a NIRspec field-of-view to be able to register the position of the telescope with the required accuracy with respect to the grid of reference targets? 
\item Are there Galactic coordinates where the NIRspec Target Acquisition integrations need to be more than the baseline duration in order to acquire and centroid enough reference sources? 
% => analyses conducted using catalogs + luminosity functions and number densities.
\item Are the positions of the scientific targets with respect to the reference stars known with a high-enough accuracy? 
% => properties of the pre-imaging data
% \item Can we use a single WFC3 field to act as a pre-imaging reference image for NIRspec observations?
\end{enumerate}

We explore specifically the edge-case where only pre-imaging with a single WFC3 field is available. It is conceivable that a few, extremely high-value targets have been identified in a single HST/WFC3 image and it might be considered worthwhile to point NIRspect in MOS mode on just one WFC3 field. Can one forgo NIRcam pre-imaging without incurring a penalty in image registration and just use the single WFC3 field?

One might consider M-type brown dwarfs specifically for several reasons. First and foremost, these are the most ubiquitous objects in the disk and halo of our Milky Way and in the near-infrared bright enough to serve as reference stars. Secondly, these have proven to be easily identified in the near-infrared imaging surveys searching for high-redshift galaxies. And thirdly, they are also bright in the mid-infrared opening up the possibility for these results to be relevant to JWST/MIRI pointing planning as well. 

\begin{figure}[htbp]
\begin{center}
	\includegraphics[width=0.5\textwidth]{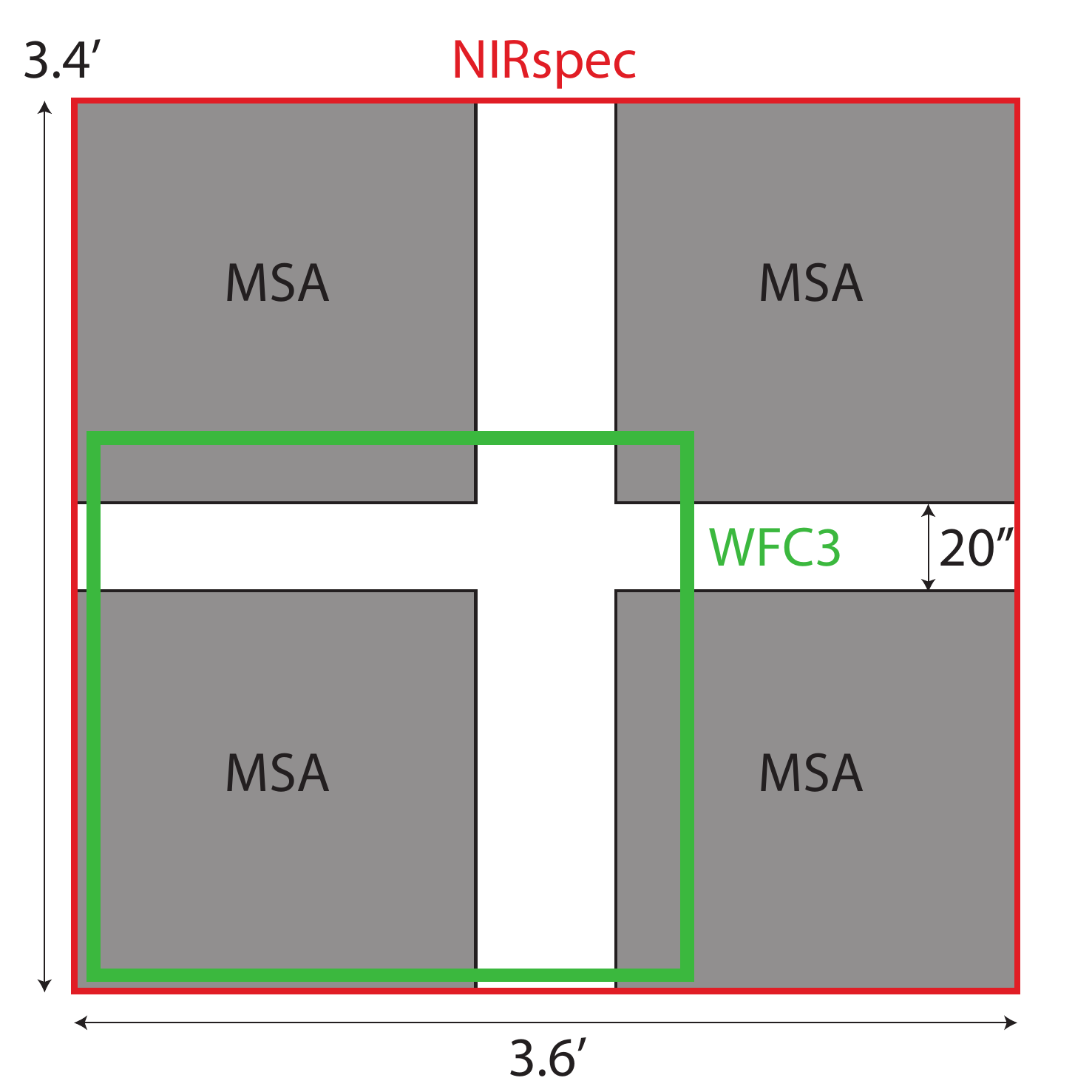}
\caption{The relative field-of-view of the WFC3/IR camera onboard the Hubble Space Telescope and the NIRspec multi-shutter array Multi-object spectroscopic observation mode. A single pointing with WFC3 does not cover all four MSA quadrants of the NIRspec instrument. }
	\label{f:fov}
\end{center}
\end{figure}

In this Research Note, we explore these questions using the Brightest of Reionizing Galaxies \citep[BoRG,][]{Trenti11} HST/WFC3 pure-parallel program. There are of course other ways to attack this problem \citep[e.g., by using the Besancon models of the Milky Way,][\url{http://model.obs-besancon.fr/}]{Robin03} but the BoRG data represent a reasonable approximation of the typical extra-Galactic field one can expect to observe with JWST/NIRspec: 
(a) the BoRG fields span a range of Galactic coordinates, 
(b) the images are in the near-infrared, 
(c) similar in pixel scale and central wavelength to the NIRspec pre-image if not in field-of-view, 
(d) the target science was the identification of extremely high-redshift galaxies ($z\sim8$ Y-band dropouts), very similar to JWST banner science, and lastly (e) an extensive catalog of identified Galactic dwarf stars based on this data is already available \citep{Holwerda14,van-Vledder16}. 

In this work, we assume the absolute on-sky WFC3 astrometry does not matter, i.e. the uncertainty in the registration of the WFC3 on the sky is not important as we only need the relative position of MSA targets and Galactic stars. Typical absolute uncertainty in HST targeting is $\sim0\farcs1$ due to reference star positional uncertainty and one is unlikely to miss reference stars or target objects as a result.

We note here that of the two acquisition filters ({\em F140X} and {\em F110W}), the scenarios considered here are closest to the use of the {\em F110W} filter. The much wider {\em F140X} filter will perform much better as it includes the I-, J- and H-band. Brown and red dwarfs are much brighter at the longer wavelengths, allowing for much shorter exposures. In a scenario where WFC3 {\em F125W} pre-imaging is used, the {\em F110W} filter is more appropriate as it will result in similar relative luminosities for the reference (and target) objects.

We aim to evaluate three metrics of accuracy: (1) the calculated astrometric accuracy for a single MSA quadrant, (2) the same for the full NIRspec MSA array, and (3) the nearest reference star to a target source, all as a function of TA image depth.

% Specifically, this note aims at providing an evaluation of the total accuracy, with which a givenNIRSpec high-redshift target with accurately known sky coordinates can be placed within a  given preselected 200 $\mu$as-wide MSA microshutter slitlet. 

This Research Note is organized as follows: 
\S \ref{s:borg} briefly describes the BoRG[z8] field data, 
\S \ref{s:cat} describes how the source catalogs were generated,
\S \ref{s:stars} gives the outline how stars are selected from these catalogs similar to \citep{Holwerda14},
\S \ref{s:z8gal} briefly describes the candidate $z\sim8$ galaxy catalog from \citep{Schmidt14},
\S \ref{s:point} addresses each of the above questions, and 
\S \ref{s:concl} lists our conclusions.

%
%\begin{figure}[htbp]
%\begin{center}
%	\includegraphics[width=0.5\textwidth]{./Figures/holwerda_f2.pdf}
%	\caption{The position of the \Nfields~ unique WFC3 Pure Parallel Fields in Galactic coordinates (red squares). The sampling of the Milky Way disk is reasonably uniform but Northern targets were preferred (because more QSOs are known from SDSS).}
%	\label{f:fields}
%\end{center}
%\end{figure}

\begin{figure*}
\begin{center}
	\includegraphics[width=\textwidth]{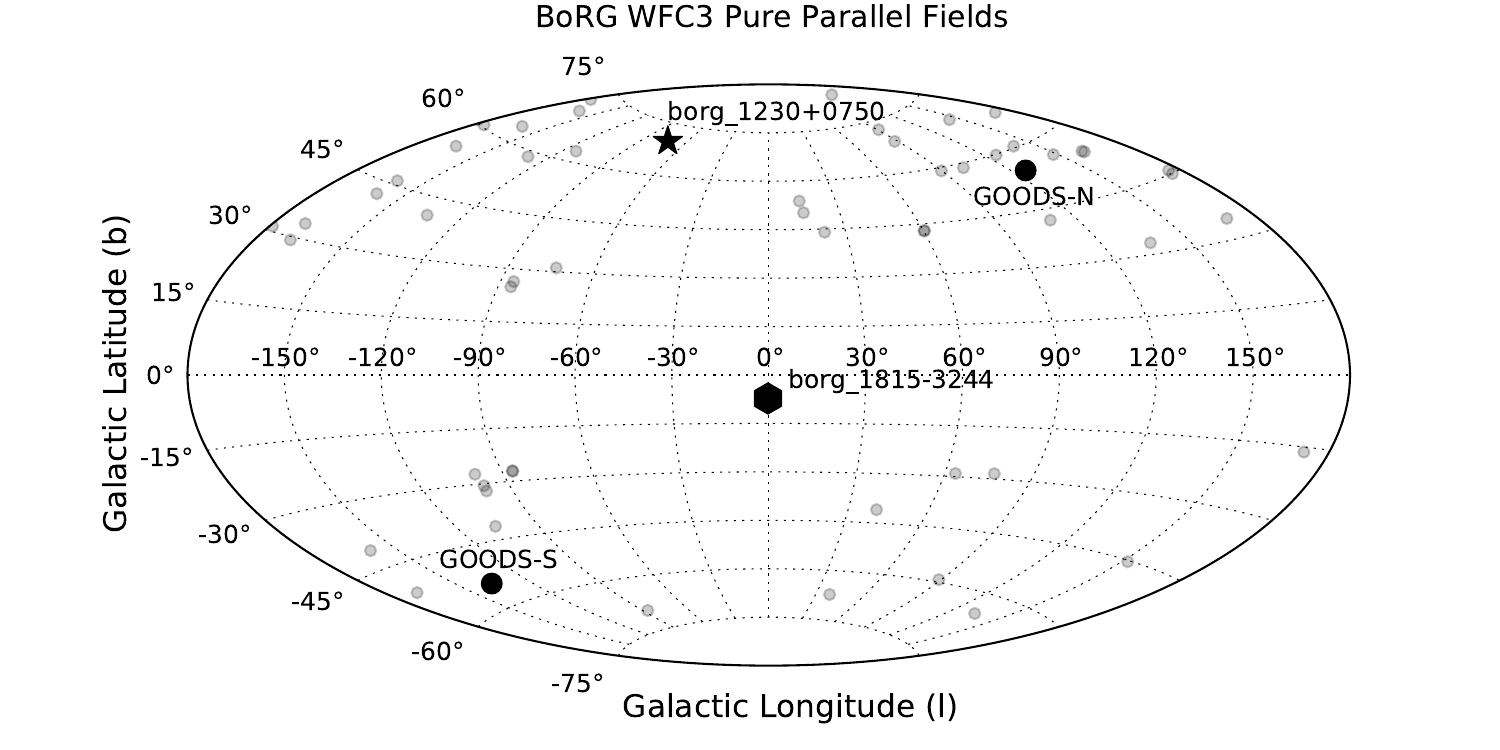}
	\caption{The position of the \Nfields~ unique WFC3 Pure Parallel Fields in Galactic coordinates (light gray points) that we utilize from the BoRG[z8] program. The sampling of the Milky Way disk is reasonably uniform but Northern targets were preferred (likely because more QSOs are known from SDSS). 
	Two BoRG[z8] fields are notably over-dense in stars, borg\_1230+0750 and borg\_1815-3244.  
	% Two BoRG are notably over-dense in stars, borg\_1230+0750 and borg\_1815-3244. 
	The former is right on the Sagittarius stream and the latter close to the Bulge. Both are omitted in this analysis.}
	\label{f:fields}
\end{center}
\end{figure*}

\begin{figure}[htbp]
\begin{center}
	\includegraphics[width=0.5\textwidth]{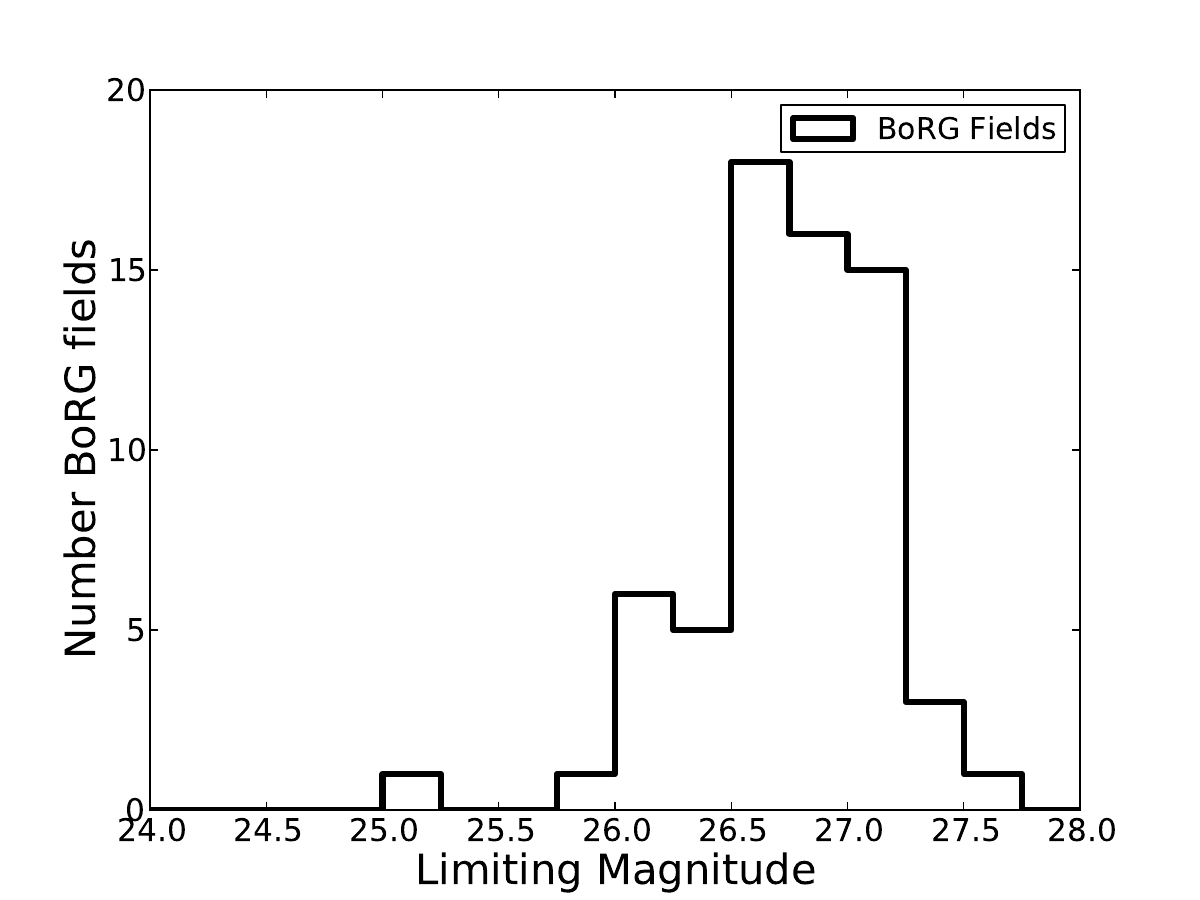}
\caption{The {\em F125W}-band magnitude limits for the BoRG[z8] fields (from \citet{Bradley12} we consider in the present study.
Photometric detections are reliable down to 26-27 magnitude.}
	\label{f:limmag}
\end{center}
\end{figure}

\section{BoRG[z8] Survey Data}
\label{s:borg}

For the pointing estimates, we use the WFC3 data from the BoRG[z8] \citep[Brightest of Reionizing Galaxies, HST GO/PAR-11700,][]{Trenti11, Bradley12, Schmidt14} survey\footnote{This is the Pure-Parallel search for $z\sim8$ search. The current BoRG[z9-10] survey primarily searches for $z\sim9-10$ sources \citep{Calvi16}.}. The BoRG[z8] observations are HST/WFC3 conducted in pure-parallel with HST pointing to a primary spectroscopic target with the Cosmic Origin Spectrograph (typically a high-z quasar at high Galactic latitude). The limitations for such observations are primarily that no dithering strategy can be used (final images are  on a larger pixel scale) and total exposure times are dictated by the primary program. The BoRG[z8] program's initial aims are to obtain as many pointings as practical using four WFC3 filters ({\em F606W, F098M, F125W}, and {\em F160W}). 

Because lines of sight are independent and well separated on the sky (mostly at high Galactic latitudes, $|b| > 20^\circ$, see Figure \ref{f:fields}), the BoRG[z8] survey both samples the Milky Way disk away from the 
plane better than single sight-lines, e.g., the GOODS \citep{goods} or CANDELS \citep{koekemoer11,Grogin11} fields, and is minimally affected by field-to-field (cosmic) variance \citep{Trenti08}. 

However, BoRG[z8] is an optimized search for $z\sim8$ galaxies through the Lyman Break technique \citep[Y-band dropout,][]{Steidel96} and not brown and red dwarf stars in the Milky Way but has proven near ideal for M- and L-dwarf searches \citep{Ryan11,Holwerda14,van-Vledder16}.
%Given the near-random pointing nature of the pure-parallel HST program (higher Galactic Latitude objects are preferred for COS targets), the BoRG[z8] fields are minimally affected by field-to-field (cosmic) variance \citep{Trenti08}. 
The BoRG[z8] data-set used here is the third data-release of \Nfields~ WFC3 fields for a total of approximately \borgarea\ arcmin$^2$ as described in detail by \citet{Trenti11}, \citet{Bradley12} and \citet{Schmidt14}, including some reprocessed fields from another pure-parallel program with similar science goals  \citep[Hubble Infrared Pure Parallel Imaging Extragalactic Survey, HIPPIES, HST GO/PAR-11702,][]{Yan10b}.

\begin{figure}[htbp]
\begin{center}
	\includegraphics[width=0.5\textwidth]{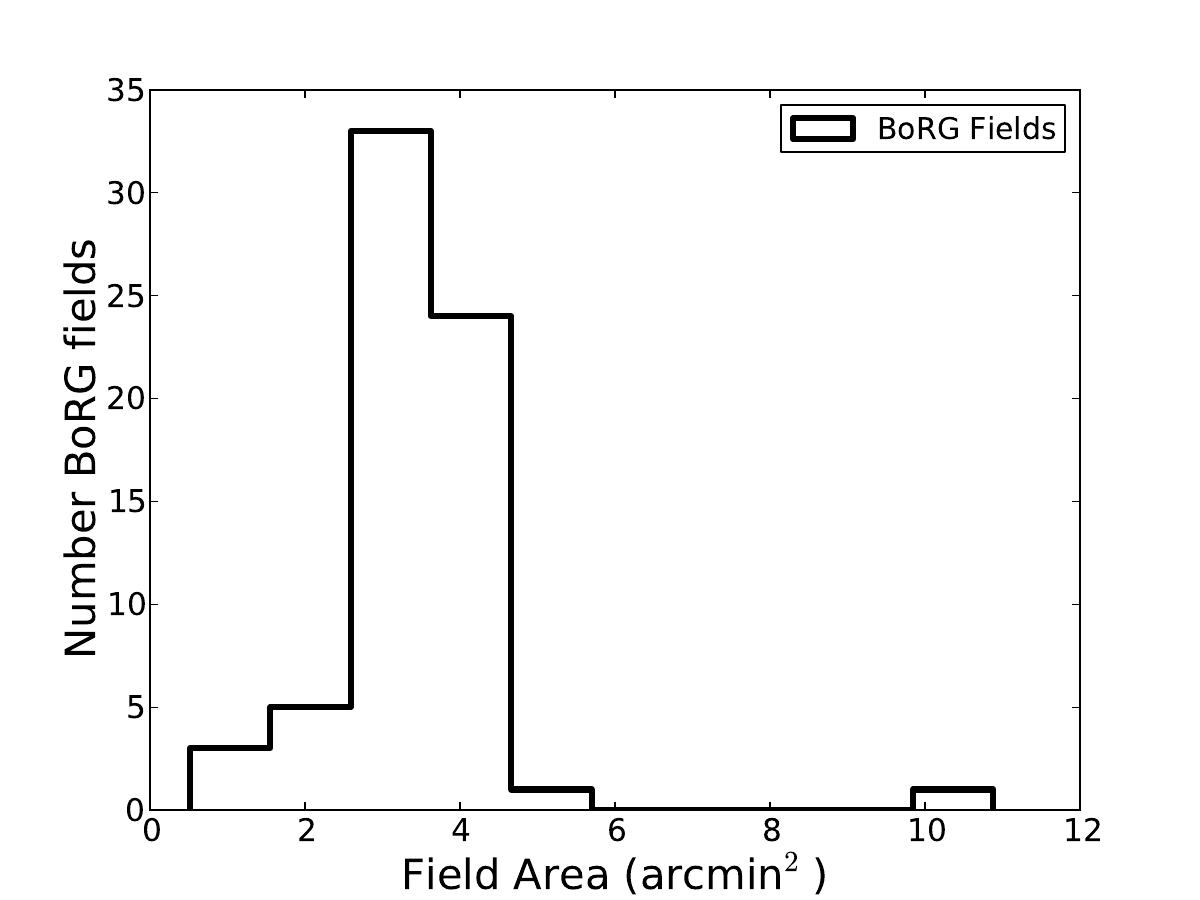}
\caption{The distribution of area covered by the BoRG[z8] pure-parallel fields. Compared with the $\sim$12 arcmin$^2$ of the NIRspec MSA field-of-view, these fields still fall short to provide full coverage of the NIRspec patrol area for MOS observations (see Figure \ref{f:fov}).}
% these still fall short to provide full coverage of the NIRspec patrol area for MOS observations (see Figure \ref{f:fov}).
	\label{f:histborgarea}
\end{center}
\end{figure}

We use the WFC3 data-products generated by the BoRG[z8] team \citep[see for details][]{Bradley12, Schmidt14}. 
This is a standard multi-drizzle reduction of these undithered WFC3 data with Laplacian edge detection \citep{van-Dokkum01} 
to the individual FLT files to mitigate detector hot pixels and cosmic rays. Stellar objects are not affected by this filtering.
Therefore, this catalog is uniquely positioned to set constraints on the number density of unresolved sources, be it Milky Way stars or $z\sim8$ galaxies. The BoRG[z8] survey pure-parallel nature means the limiting depth and field area varies due to the requirements of the primary observation (integration time, roll angles). The field area is defined as the part of the image suitable for the detection of $z\sim8$ galaxies, the nominal target for the BoRG[z8] survey \citep[removing bad pixels, chip gaps, and those parts of the field without Y-band imaging][]{Bradley12}.
The distribution of limiting magnitudes and usable on-sky area are shown in Figures \ref{f:limmag} and \ref{f:histborgarea} respectively.

\section{Catalog Generation}
\label{s:cat}

To construct the catalogs for the BoRG[z8] Fields, we ran {\sc sextractor} \citep{SE,seman} with identical settings as \citet{Trenti11} but set to include {\sc mu\_max}, {\sc flux\_radius} ($r_{50}$) and other morphological information in the output catalogs to aid in the identification of stars and the uncertainty in their position.% (Table \ref{t:se}). 
% NB DO I NEED TO LIST THE SEXTRACOR SETTINGS?
% Tables \ref{t:se:in} and \ref{t:se:out} lists the typical input parameters and the output.
% detection
Source detection was done in the {\em F125W} image, with the other filters run in dual-image mode. The multi-drizzle weight files were used as RMS maps  \citep[once normalized, following the prescription in][{\sc WEIGHT\_TYPE=MAP\_RMS}]{Casertano00}, but scaled appropriately to reflect the noise correlation introduced, typically a factor $\sim$1.1 or less in each filter (see \citet{Trenti11}, \citet{Bradley12} and \citet{Schmidt14} for details). 
% A requirement of S/N $> 8$ in {\em F125W} and S/N $> 3$ in {\em F160W} was used for the final catalog. 
The photometric zero-points are from \citet{Dressel10, Windhorst11}: {\em F606W}: 26.08, {\em F600LP}: 25.85, {\em F098W}: 25.68, {\em F125W}: 26.25, and {\em F160W}: 25.96 with Galactic $A_V$ corrections derived for each individual field. These settings resulted in \nborg\ objects in the BoRG[z8] survey.

\begin{figure}
\begin{center}
\includegraphics[width=0.5\textwidth]{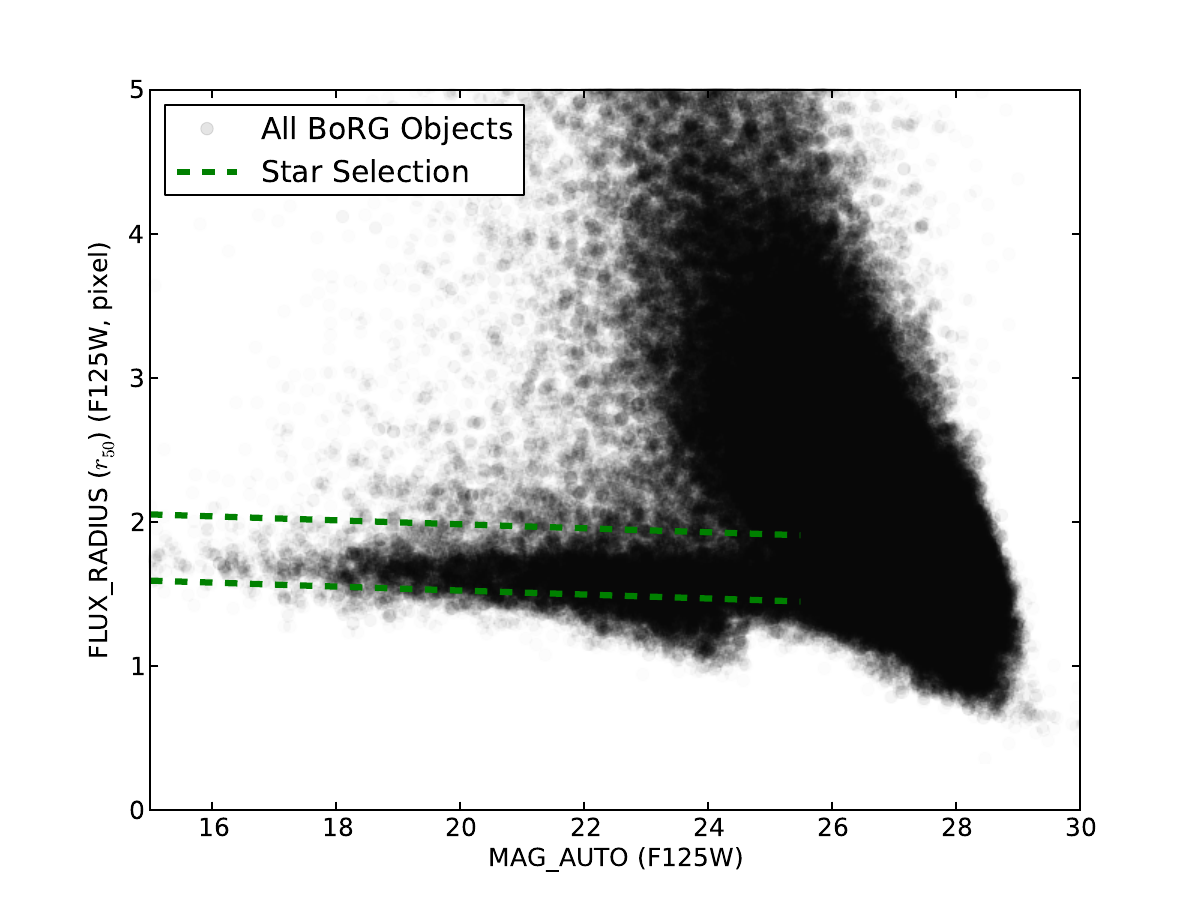}
\caption{The morphological selection criterion we use to select stars from the BoRG[z8] fields to use for astrometric alignment sources.   Our selection criteria (green dashed line, equation \ref{eq:r50}) are based on the effective radius ({\sc FLUX\_RADIUS} or $r_{50}$) and the apparent magnitude in {\em F125W} derived from bona-fide (spectroscopically confirmed) dwarfs in the CANDELS field.}
\label{f:r50}
\end{center}
\end{figure}

\begin{figure}
\begin{center}
\includegraphics[width=0.5\textwidth]{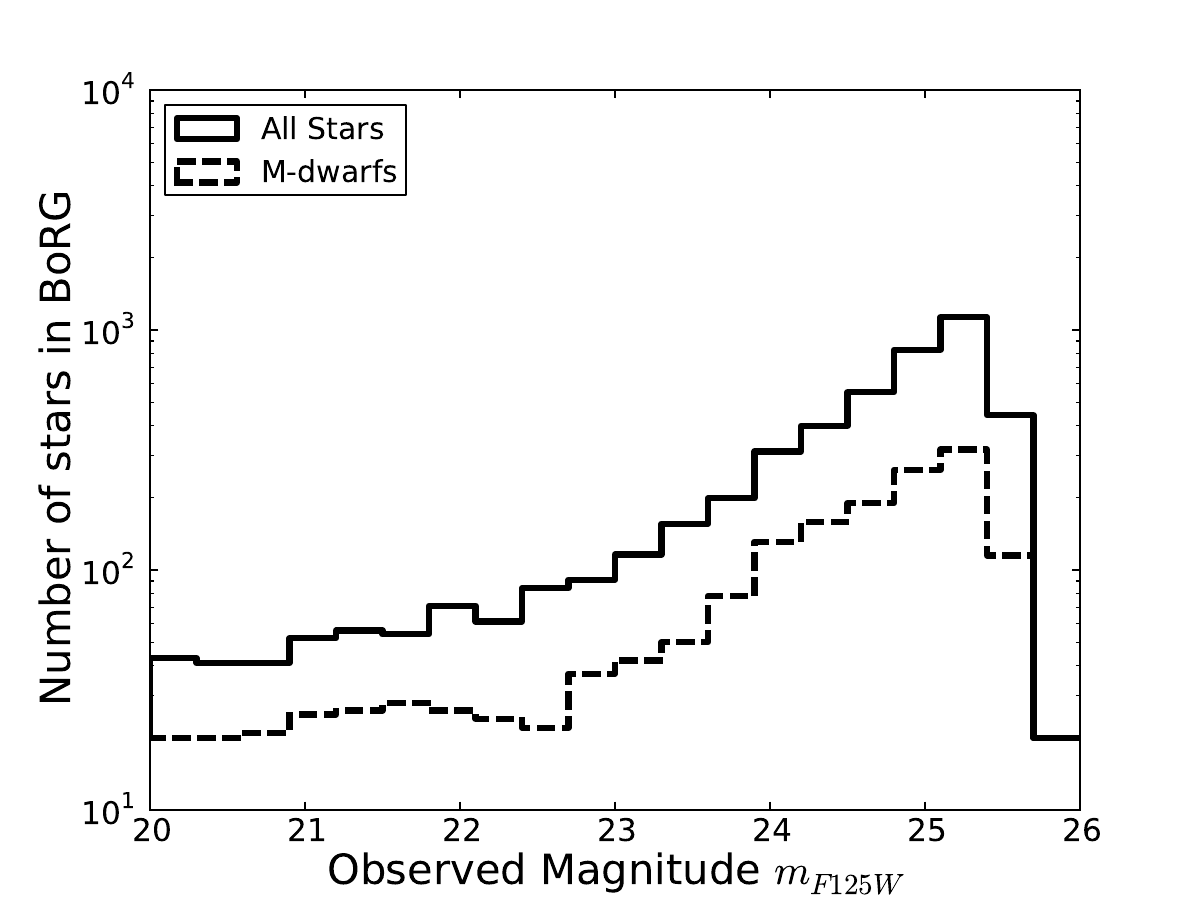}
\caption{The apparent {\em F125W} luminosity function of stars in all the BoRG[z8] fields combined. The solid line is all stellar objects, selected morphologically in the {\em F125W} source extractor catalog (Figure \ref{f:r50}) and the dashed line are those stars which satisfy the Y-J, J-H color-color selection for M-type stars.}
\label{f:LF125}
\end{center}
\end{figure}

\section{Morphological Selection of Stars}
\label{s:stars}

The stars were selected using an optimal morphological identification, reported in \citet{Holwerda14}\footnote{Continuing work on star/galaxy separation in crowded fields started in \citet{mythesis,Holwerda05a,GHOSTS}.}, calibrated by the morphology of bona-fide M-dwarf stars in the CANDELS v0.5 mosaic on GOODS-North. Briefly, we selected those stars identified by \citet{Pirzkal09} in the PEARS {\em grism} survey of GOODS-North as M-dwarfs based on a fit to their spectra. We then plotted these in a luminosity versus effective radius plot from the source extractor catalog of the {\em F125W} CANDELS image, resampled to BoRG[z8] pixel scales. Their position served to map the selection criterion below \citep[see][for a complete discussion]{Holwerda14}.

Similar to \citet{Ryan11}, we found that the effective radius ({\sc FLUX\_RADIUS} or $r_{50}$) of objects is an excellent identifier of unresolved, stellar objects (Figure \ref{f:r50}). The BoRG[z8] color-color information is primarily optimized to identify $z\sim8$ galaxy candidates but can also be used to identify M, L and T-dwarfs.
The stellar selection consists of two steps: a morphological selection and subsequent stellar typing using the colors. Based on the {\em F125W} filter information, we identify stars through the following relation:
\begin{equation} 
\label{eq:r50}
| r_{50} -0.014 \times MAG\_AUTO-2.03 | < 0.23.
\end{equation} 
In \citet{Holwerda14}, we limited the applicable range to $m_{F125W}<24$ mag (their Figure 8) to cleanly select and subtype M-dwarfs but if one only needs to identify stars, this relation can be extended to 25.5 mag. Using the above relation a total of \nstars\ stars can be identified in the BoRG[z8] fields. Of this total of \nstars\ stars, \Mdwarfs\ satisfy the M-type and \Ldwarfs\ L-type dwarf color-color selection from \citet{Ryan11},  \citet{Holwerda14} and \citet{van-Vledder16}.
Figure \ref{f:LF125} shows the apparent luminosity distribution of all the stars as well as the M-dwarfs.

\section{Fiducial NIRspec Targets; $z\sim8$ Galaxies}
\label{s:z8gal}

\citet{Bradley12} present the first sample of candidate $z\sim8$ galaxies in the BoRG[z8] fields and \citet{Schmidt14} followed with a supplemental catalog, bringing the total of $z\sim8$ Y-band dropouts to \nzeightgal. 
We use the combined catalog as the fiducial high-value target list in section \ref{s:neareststar}. In reality, the number density of $z\sim8$ galaxies in a single WFC3 pointing like in the BoRG[z8] survey is generally too low to justify NIRSspec MSA observations and one would more likely go for a much simpler single-slit observations, although other lower redshift targets might be interesting for spectroscopic follow-up as well. We use this science case an one example of sources identified from HST imaging in single WFC3 fields that might be of interest for JWST spectroscopy. 
Our assumption is that all these candidate high-redshift galaxies are bona-fide, even though \citet{Bradley12} divide the sample between $8\sigma$ and $5\sigma$ confidence-level detections and \citet{Schmidt14} note that not every $z\sim8$ galaxy is confirmed in the single re-observed field.
Actual NIRspec targets will be similarly selected based on their colors or possible a full spectral energy distribution fit to determine their photometric redshift (photo-z) and will also suffer from false detections. 
 For the purpose of this Research Note, as the fiducial high-value NIRspec spectroscopy targets, the BoRG[z8] $z\sim8$ candidates and their positions will suffice to explore the relation to nearby guide stars.

\begin{table*}[htdp]
\caption{\label{t:specs}The image characteristics of the WFC3 undithered exposures and the NIRspec pre-spectroscopy image.}
\begin{center}
\begin{tabular}{l l l l}
				& WFC3			& NIRspec 		& Unit						\\
\hline
\hline
FOV				& 2.05 $\times$ 2.27	& 3.4 $\times$ 3.6	& arcmin $\times$ arcmin			\\
pixelsize 			& 0\farcs08		& 0\farcs1			& arcsec						\\
central wavelength	& 1.6				& 0.6-5			& $\mu$m.					\\
\hline
\end{tabular}
\end{center}
\end{table*}%

\section{NIRspec Target Acquisition}
\label{s:point}

The WFC3 observations and the NIRspec TA image are close in general characteristics (Table \ref{t:specs}) but the field-of-view of a single WFC3 exposure does not cover the whole of the NIRspec one (Figure \ref{f:fov}). 
% A large fraction of the guide stars will be M-dwarfs in the Galactic Disk and Halo (Figures \ref{f:LF125}). 
Starting from the $z\sim8$ targets and the above morphological selection of stars, we aim to answer the following questions:
\begin{itemize}
\item What is the reference stars surface density?
\item Are there Galactic Coordinates where there are much more or much less reference stars? 
\item What is the typical uncertainty in a NIRspec TA image position? 
\item Will we know the position of the scientific targets with respect to the reference stars with a high-enough accuracy?
\end{itemize}

\begin{figure}
\begin{center}
\includegraphics[width=0.5\textwidth]{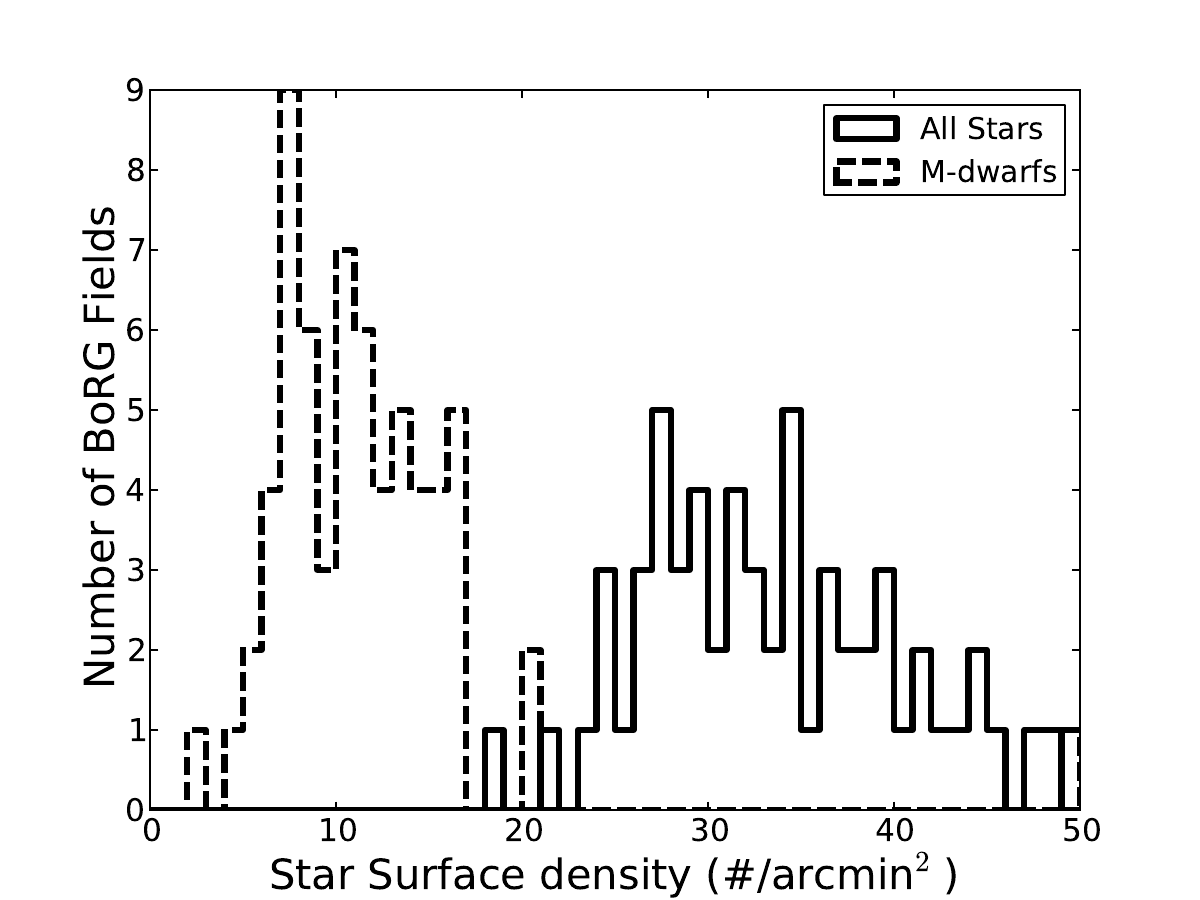}
\caption{The distribution of number densities of $m_{F125W} < 25.5$ stars, of any type and M-dwarfs only in the BoRG[z8] fields. }
\label{f:histNstar}
\end{center}
\end{figure}

\begin{figure}
\begin{center}
\includegraphics[width=0.5\textwidth]{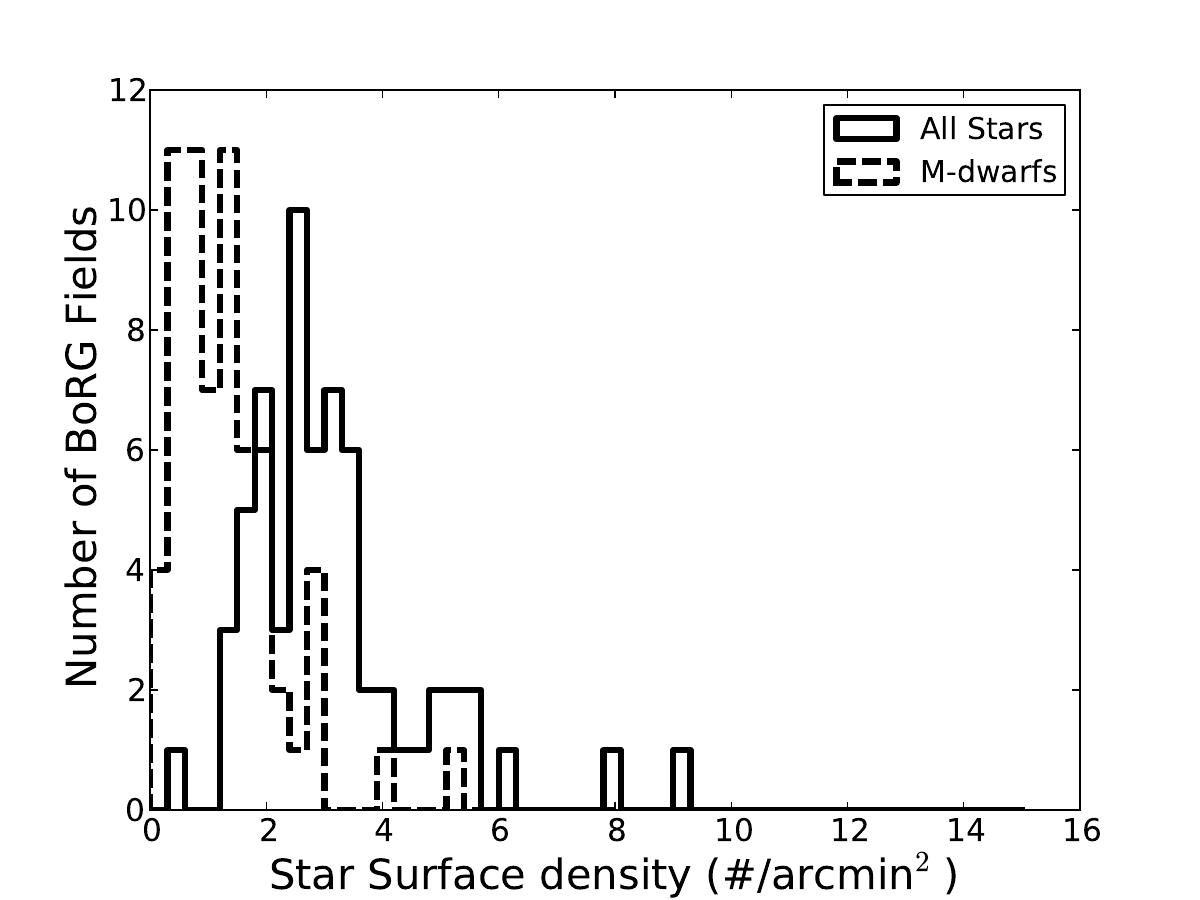}
\caption{The distribution of number densities of $m_{F125W} < 24$ stars, of any type and M-dwarfs in the BoRG[z8] fields. }
\label{f:histNbrightstar}
\end{center}
\end{figure}

\subsection{NIRspec Registration Requirements}

The NIRspec TA image happens with all MSA slits open. A short exposure is taken in one of the TA filters ($F140X$ or $F110W$) and then the MSA is moved half a slit in both the spatial and dispersion axes and the short exposure is repeated. In order to have the onboard centeroiding software accurately determine the sky position of the NIRspec MSA, several stars need to be already selected from either a HST or NIRcam pre-image. Part of our motivation is to consider if a single WFC3 image or a mosaic would have enough stars in it to register NIRspec already and no NIRcam image will need to be made.

NIRspec team simulations \cite[see][]{Regan05,Beck09,Beck15,Karakla15,Ubeda15}  have shown that between 8-20 stars are needed for the centroiding to be successful enough for a solution with 10\% of the MSA slit width (20 milli-arcseconds). However, the stars considered were over the NIRspec full area and typically brighter (14-18 AB mag) than considered here.

% \subsubsection{Reasons for deeper NIRspec TA imaging}

There are two good reasons to go deeper on the TA imaging and obtain fainter reference stars: (1) the fainter stars are more likely to be further in the disk or halo of the Galaxy, making their proper motions much less of a concern and (2) many more fainter stars are available per area of sky, improving the chances of centroiding success.

For example, a typical star in the disk travels with $\sim$100 km/s and at a distance of $<$5 kpc \citep[over 10x the scale-height of the Galactic disk][]{van-Vledder16} this translates to 4-5 milli-arcseconds in a 10-year time period (the worst case scenario using the oldest WFC3 data to point NIRspec). Adding more distant stars will mitigate the random proper motions of the reference stars in the field.

\subsection{What is the surface density of reference stars?}
\label{s:fov}

The luminosity functions in Figure \ref{f:LF125} already point to a wealth of stars on average in each of the BoRG[z8] fields.  
Figure \ref{f:histNstar} shows the distribution of the stellar density in all the BoRG[z8] fields. BoRG[z8] fields are preselected to be higher ($|b|>10^\circ$) % Galactic Latitude but many Galactic stars are typically present. 
Galactic Latitude but many Galactic stars are still typically present. 
Brighter than 25.5 magnitude in $F125W$, even the sparsest field has 20 stars of any type in it. 
% Restricting oneself to M-dwarfs even will practically ensure enough stars to register an image (the minimum number for that purpose is three), with the exception of a single field. If the requirement is more stringent ($m_{F125W}<24$ and M-dwarfs alone), only tree fields would fail to have less than three stars per arcmin$^2$!
We can therefore safely conclude that NIRspec fields at high Galactic latitude are very likely to have enough stars in the field-of-view to register the image if they are deep enough. Whether this ensemble of reference stars can successfully astrometrically register NIRsec through a TA image, is explored in section \ref{s:wfc3pos}.

\begin{figure*}
\begin{center}
\includegraphics[width=0.5\textwidth]{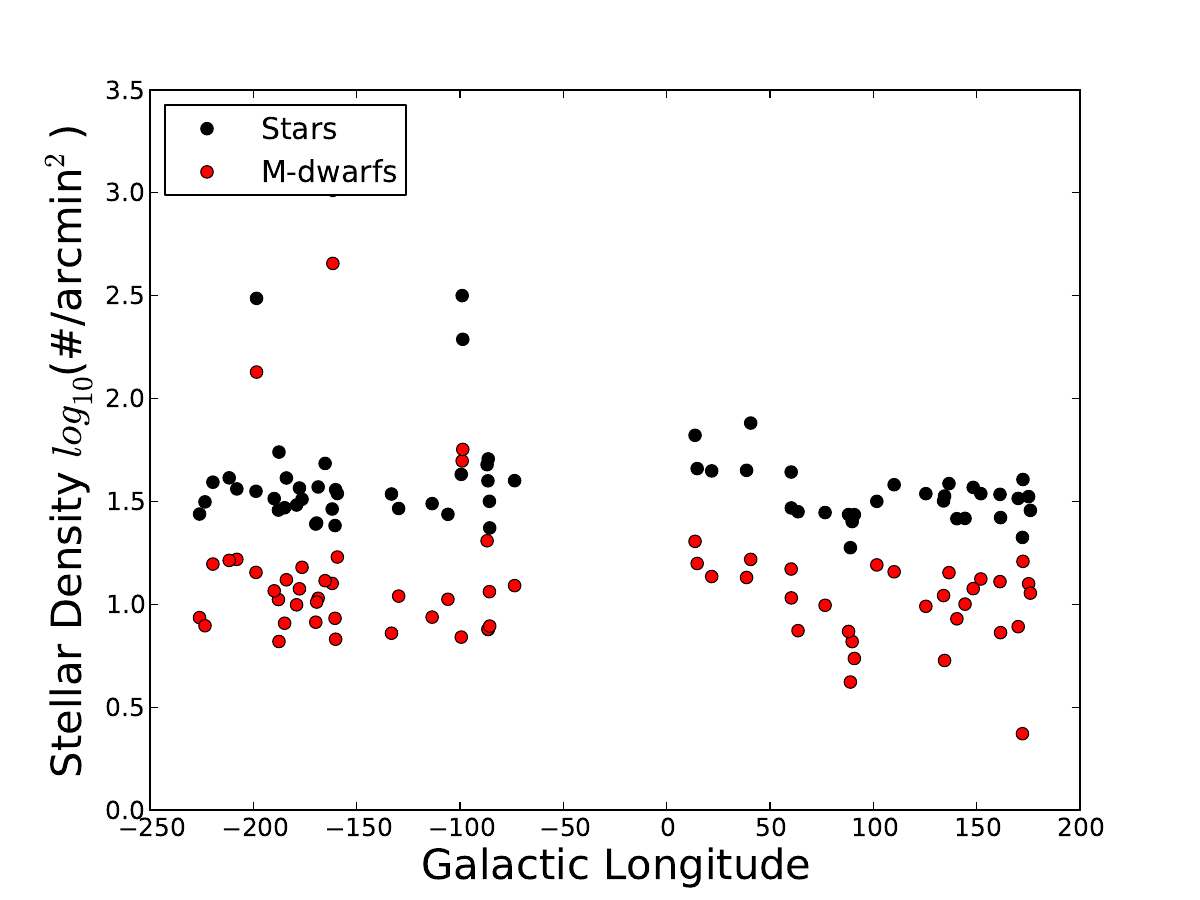}
\includegraphics[width=0.5\textwidth]{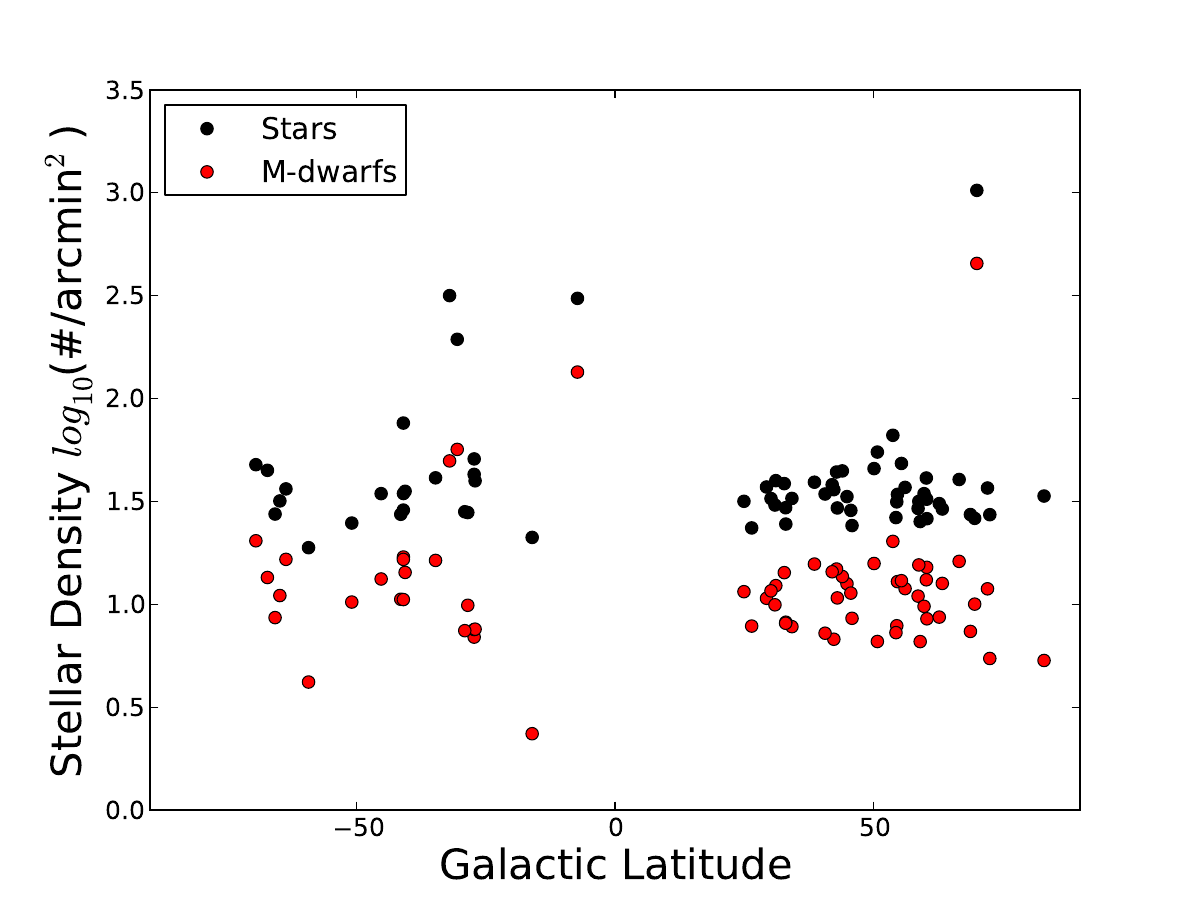}
\caption{Stellar Surface density as a function of Galactic Longitude (left) and latitude (right). }
\label{f:NstarGal}
\end{center}
\end{figure*}

\subsection{Are there Galactic Coordinates where there are not enough stars?}
\label{s:surfacedensity}

This question remains if there are specific Galactic Coordinates where there are fewer reference stars than others. One can reasonably expect so with in-the-plane fields being much more crowded
and a dependence of scale-height of dwarfs on the latitude for example. Figure \ref{f:NstarGal} shows the density of $m_{F125W} < 25.5$ stars as a function of the Galactic coordinates. The effect of the Bulge can be
seen in a few fields but overall, the field-to-field variation is much greater than any trend with Galactic coordinates. 
Stellar streams --such as the Sagittarius stream identified in M-dwarfs in \citet{Holwerda14} -- have a greater effect on the number of dwarfs in a given field than for example the Bulge. For typical extra-Galactic targets ($|b| > 10^\circ$), there are at least this number of reference stars available, regardless of position on the sky: $log_{10}(\#/arcmin^2) \sim1.5$. If one restricts oneself to M-dwarfs, half a dex less.

\begin{figure}
\begin{center}
\includegraphics[width=0.5\textwidth]{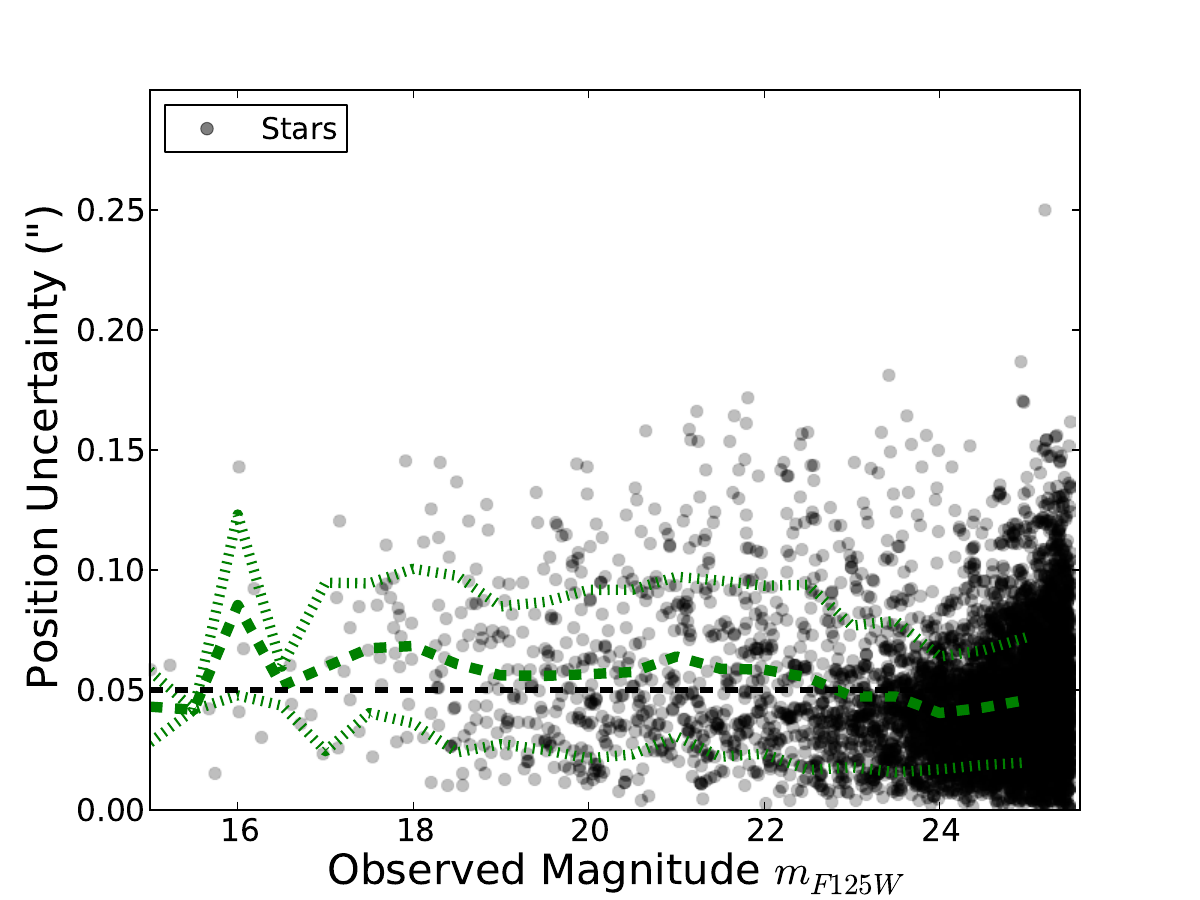}
\caption{The {\sc source extractor} estimate of the position error in stars as a function of their apparent magnitude. The green dashes and dotted lines are the mean and standard deviation in the stellar astrometry. The black dashed line is the astrometric accuracy required for individual stars in NIRspec pre-spectroscopy imaging.}
\label{f:starposerr}
\end{center}
\end{figure}

\subsection{What is the typical uncertainty in a NIRspec TA image (based on WFC3 imaging)?}
\label{s:wfc3pos}

One question regards what the typical uncertainty in a NIRSPEC pointing would be based on the available WFC3/IR imaging.     
To answer this question, we will have to assign an uncertainty in the position to each of the stars selected from the {\sc source extractor} catalogs. 
{\sc Source Extractor} provides a number of position outputs for every objects in the catalog and uncertainties based on the pixel-to-pixel rms \citep{SE,seman}. 
The catalogs contain two pixel positions for each object, the flux-weighted center (zeroth-order moment or barycenter; {\sc x\_image, y\_image}) and the position of the brightest pixel ({\sc x\_peak, y\_peak}). The barycenter position is a close analog to the centroiding position in the NIRspec acquisition image.
Unfortunately, {\sc source extractor} does not provide rms-based uncertainty estimates for the barycenter and none is physical for the discrete 
value of the peak pixel (barring a PSF fit). 

Thus, in order to estimate the uncertainty in the position of stars, we have to make an approximation and we use the two positions (barycenter and peak) for this purpose. If the position is known to within a pixel, the difference between these two can at most be 0\farcs04. If however, there is significant noise in the barycenter determination, the difference should be more for unresolved objects.
% One can argue that either represent a reasonable choice for an objects position. In case of a star, an unresolved object, these two positions should be identical. If the star's signal-to-noise is high enough, its position should be well known. If it is not, the positions may differ.
%
% One can use the difference in the two pixel positions as the uncertainty in the position of a given star:
We adopt this approximation of the astrometric uncertainty using {\sc source extractor} positions for stars:
\begin{equation}
\label{eq:starpos}
\sigma_{Star} = {1 \over F} \sqrt{(X_{IMAGE}-X_{PEAK})^2 + (Y_{IMAGE}-Y_{PEAK})^2  },
\end{equation}
\noindent where $F$ is the star's flux. The estimated uncertainty is shown in Figure \ref{f:starposerr} as a function of luminosity. 
% This estimate of astrometric uncertainty in individual stars is shown as a function of apparent magnitude ($m_{F125W}$) in Figure \ref{f:starposerr}. 
The limiting magnitude of the BoRG[z8] fields is $\sim26.5$ (Figure \ref{f:limmag}). We adopt Figure \ref{f:starposerr} positional uncertainties as fiducial for the NIRspec centroiding algorithm uncertainty (see below). The mean stellar uncertainty in these WFC3 fields does not appear to increase with limiting magnitude (green dashed line in Figure \ref{f:starposerr}). This astrometric uncertainty does not include proper motion due to the random motion of the stars in the period between the HST/WFC3 observations and the future NIRspec observations.

% assumptions for WFC3 field accuracy
% 1. stars can be identified about 1 mag above limiting mag.
% 2. the position of dwarf stars around the z~high galaxies is already known from discovery imaging.
% 3. XY is relevant, roll angle much better constrained.
% 4. it is OK to use a WHOLE WFc3 field.
 
We need to make several assumptions to arrive at the distribution of NIRspec TA accuracy as function of image depth based on the BoRG[z8] field data: 
(1) stars can be reliably identified about 1 magnitude above the limiting magnitude by the centroiding software in the MSA pre-image (i.e., we only select stars brighter than 1 mag above the limiting magnitude, the practical limit for stellar identification in the BoRG[z8] fields), 
(2) the position of the stars, together with the target(s) is already well known from the WFC3 data on which the scheduled observation is based 
(equation \ref{eq:starpos}) and any proper motion term in the uncertainty in the stellar positions can be ignored, 
(3) the initial Guide star acquisition of JWST produces a much higher accuracy in the roll angle (position angle of the slits) than in the RA,DEC (or X,Y shift in pixels), allowing us to ignore the roll angle term in the field registration \citep{Beck09,Beck15},
(4) the absolute WFC3 registration on the sky does not need to be taken into consideration as we need to know the {\em relative} position of the stars to the target, 
and 
(5) we use {\em all} the stars in a WFC3 field to compare the field registration accuracy, even though only part of the WFC3 field-of-view will be in the MSA pre-image (Figure \ref{f:fov}). We introduce a factor 3/2 below to account for the difference on-sky between WFC3 and a MSA quadrant.
(6) Implicitly, we assume the  {\em F110W} TA filter as it is close in characteristics to the  {\em F125W} used here. There is, however, a much wider TA filter available ({\em F140X} $0.8 < \lambda < 2.0 \mu$m.), which would shorten TA exposures significantly.
% We note here that of the two acquisition filters ({\em F140X} and {\em F110W}), the scenarios considered here are closest to the use of the {\em F110W} filter. The much wider {\em F140X} filter will perform much better as it includes the I-, J- and H-band. Brown Dwarfs are much brighter at the longer wavelengths, allowing for much shorter exposures.

To estimate the uncertainty in the registration of a field's position, we take:
\begin{equation}
\label{eq:fieldpos}
\sigma_{Field} (m_{lim}) = {3 \over 2 N} \sqrt{  \Sigma_i^N (\sigma_{star}^i)^2 \times F^i \over  \Sigma_i^N F^i}
\end{equation}
for each field,where $N$ is the number of identifiable stars 1 mag above the limiting magnitude, $F^i$ the flux of that star, and $\sigma_{star}$ is the uncertainty in each star in the field. The factor $3/2$ is to account for the discrepancy in WFC3 to the MSA quadrant FOV.

The NIRspec team also estimated the MSA TA accuracy \citep[][Beck et al. {\em private communication}]{Beck15} with this expression:
\begin{equation}
\label{eq:PA} 
PA = \sqrt{ (A_{ast}^2 + 104) + \sqrt{ \left( {647.5 + A_{ast}^2 \over \sqrt{N_{stars}}} \right)^2 + 62)}+ 75 }
\end{equation}
\noindent where the PA is the NIRspec MSA Position Accuracy in milli-arcseconds, $A_{ast}$ is the positional accuracy of each of the number of stars in the field ($N_{stars}$).
This estimated positional accuracy derived from the average number of stars in the BoRG[z8] fields (but scaled up to the NIRspec MSA field-of-view) is shown as the dark gray line in Figure \ref{f:FieldPosErr}.

\begin{figure}
\begin{center}
\includegraphics[width=0.5\textwidth]{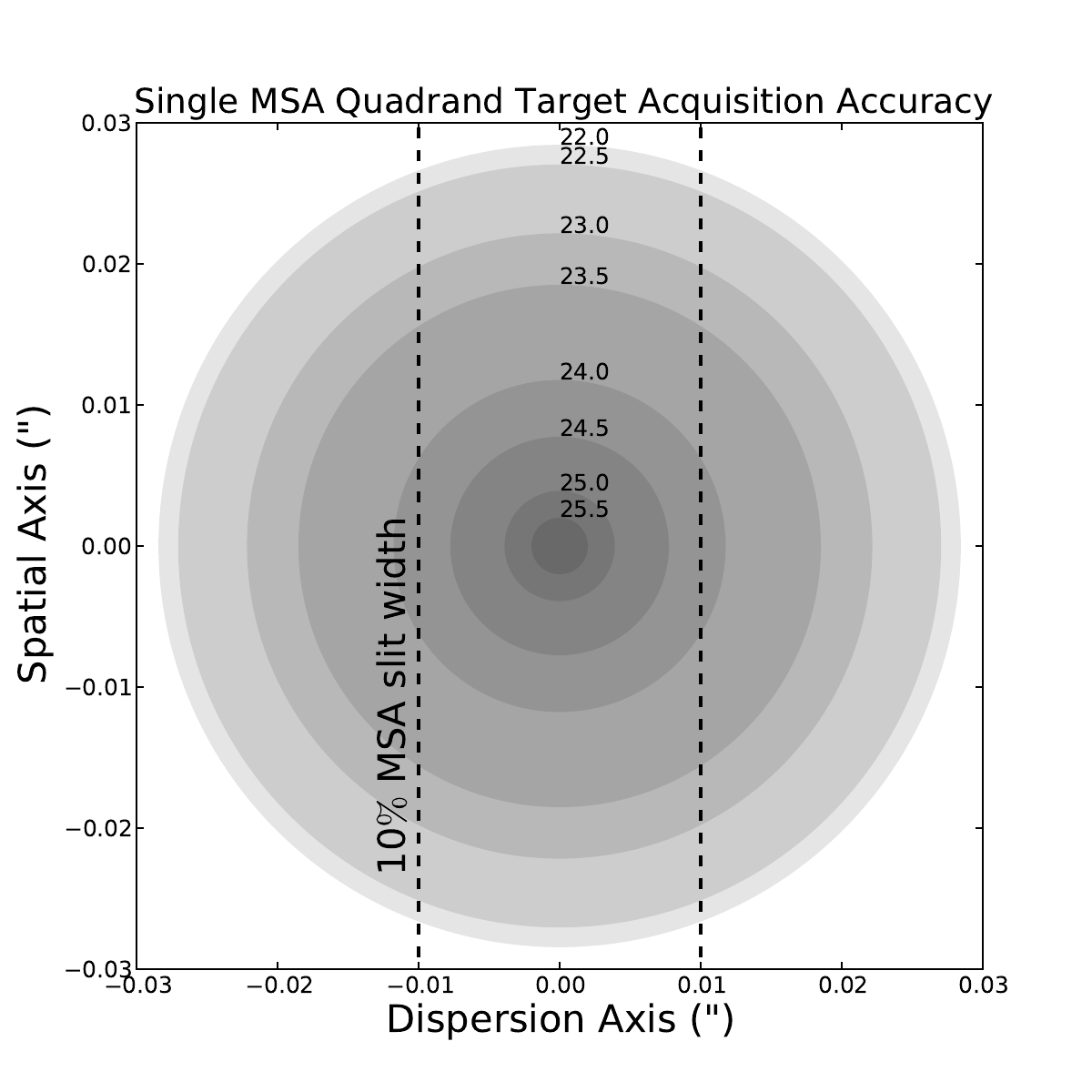}
\caption{The relation between limiting magnitude  and WFC3 field registration uncertainty, assuming the mean stellar positional uncertainty is  that of Figure \ref{f:starposerr}. The dashed line is the nominal accuracy of the astrometry required for NIRspec observations (10\% of a MSA slit). }
\label{f:FieldReg}
\end{center}
\end{figure}

\begin{figure}
\begin{center}
\includegraphics[width=0.5\textwidth]{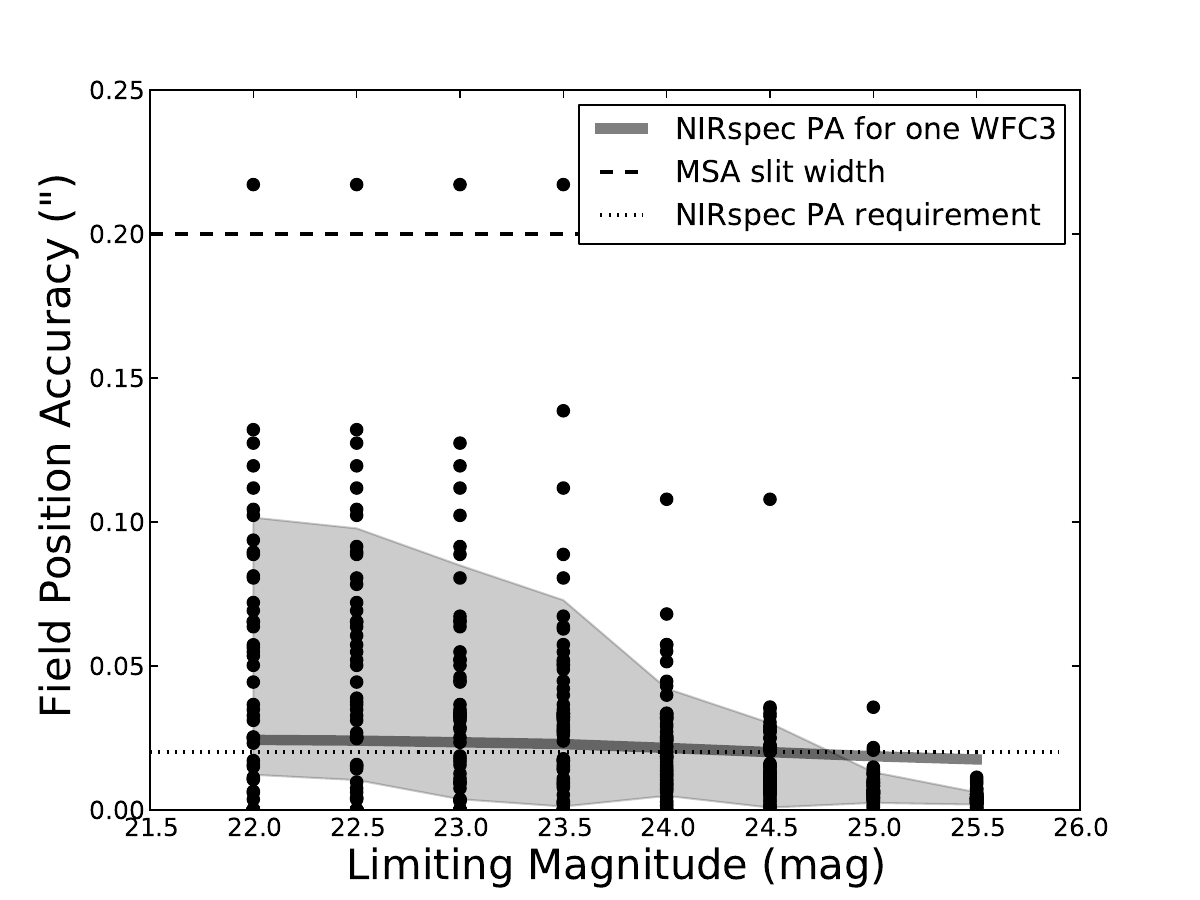}
\caption{The relation between limiting magnitude of fields and the mean and spread in TA accuracy, assuming the stellar positional uncertainty is typical that of Figure \ref{f:starposerr}. The light gray shaded area is the 1 $\sigma$ spread around the mean of the distribution. The dark gray line is the estimated MSA registration accuracy using the stars in the BoRG[z8] field (equation \ref{eq:PA}).
The dashed line is the width of the MSA slit, 10\% of which is the minimum astrometric accuracy for NIRspec observations (dotted line). }
\label{f:FieldPosErr}
\end{center}
\end{figure}

Figures \ref{f:FieldReg} and \ref{f:FieldPosErr} show the mean and the spread of NIRspec TA astrometric accuracy based on the 
BoRG[z8] WFC3 fields assuming the stellar uncertainties in Figure \ref{f:starposerr} and the expression in equation \ref{eq:fieldpos}.
Variation in individual BoRG[z8] fields in Figure \ref{f:FieldPosErr} occurs because not every BoRG[z8] field is of exact equal size (Figure \ref{f:histborgarea}) and there is natural variance in the number of available stars in each field. 

% SINGLE WFC3
Single WFC3 fields will provide positional accuracy to within 10\% MSA slit width if the HST and NIRspec pre-imaging adopt a limiting magnitude of $\sim24-24.5$, in order to select enough stars and characterize their position with enough accuracy (Figure \ref{f:FieldReg}). 
Taking the variance in the BoRG[z8] fields into account, one would need a magnitude deeper to ensure this accuracy in all cases. 

\begin{figure}
\begin{center}
\includegraphics[width=0.5\textwidth]{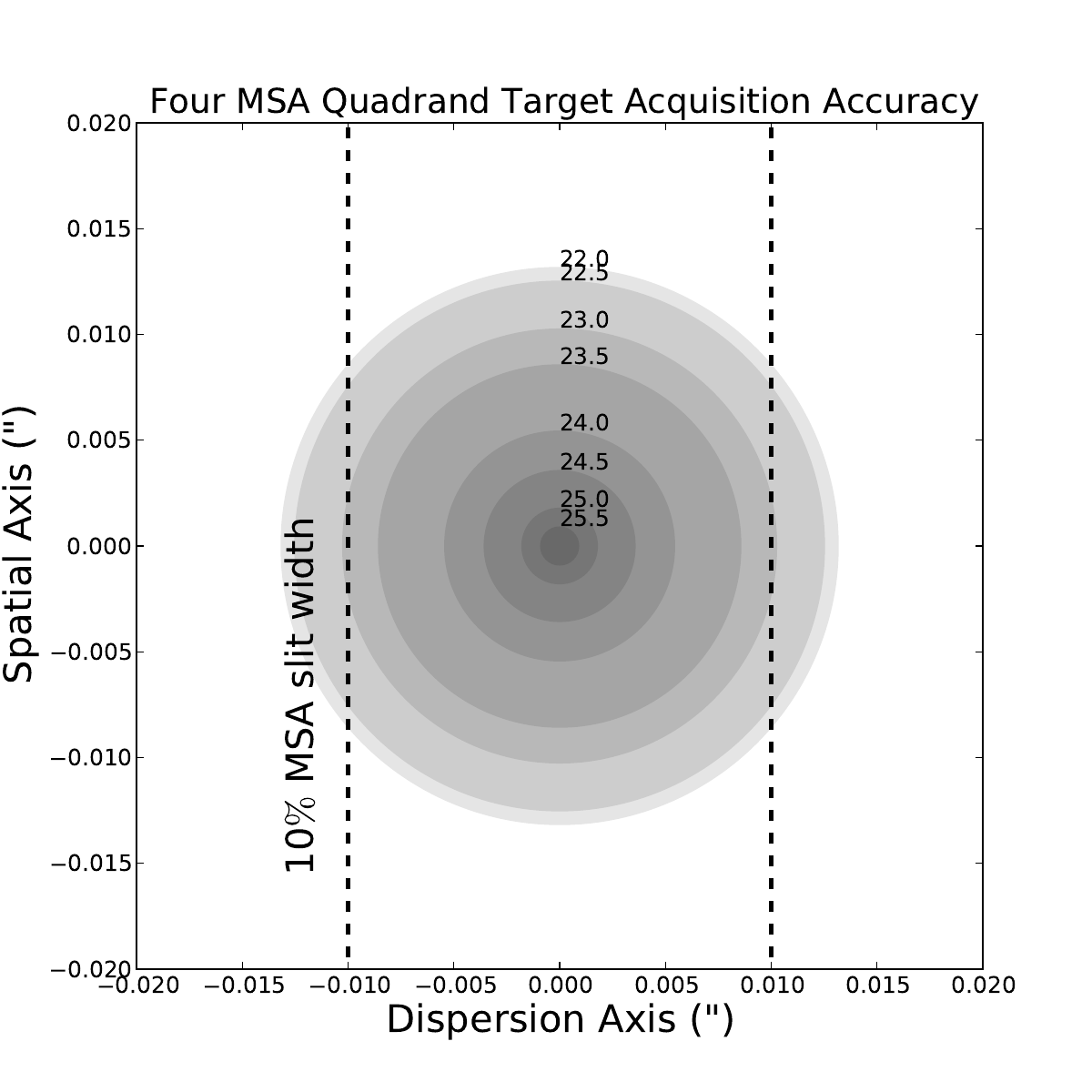}
\caption{The relation between limiting magnitude and NIRspec registration uncertainty, assuming the mean stellar positional uncertainty is typical that of Figure \ref{f:starposerr} and the full NIRspec MSA array can be used for image astrometry (see Figure \ref{f:fov}).  The dashed line is the required TA accuracy, 10\% of the MSA slit width. The difference between this Figure and Figure 11 illustrates the gains that one can
achieve when WFC3/IR imaging is available over the full $3\farcm6 \times 3\farcm6$ FOV of
NIRSPEC.   Such gains will be available over extended WFC3/IR mosaics (e.g., 
in CANDELS), but not generally over single pointing programs like BoRG[z8].
}
\label{f:MosReg}
\end{center}
\end{figure}

% WFC3 mosaic
% In the case of WFC3 mosaics (e.g., CANDELS or expanded BoRG[z8] fields), the all four MSA quadrants of NIRspec can be used for pre-spectroscopy image registration. 
In the case of WFC3 mosaics (e.g., CANDELS or expanded BoRG[z8] fields), all four MSA quadrants of NIRspec can be used for pre-spectroscopy image registration. Given the expression in equation 3, the accuracy simply scales with effective field-of view ($\sim2.16$ from WFC3 to all four MSA quadrants). Figure \ref{f:MosReg} shows the mean TA accuracy based on the BoRG[z8] fields but scaled to the full NIRspec field-of-view. A limiting magnitude of a $m_{F125W}<23$ AB suffices for TA ($\sim10$\% of the 0\farcs02 MSA slit-width). 

Thus, both a {\em single} WFC3 image as well as a mosaic can reliably function as a pre-image for NIRspec observations, provided the NIRspec TA image is of appropriate depth (24.5 and 23 AB mag in $F110W$).

\begin{figure}[htbp]
\begin{center}
	\includegraphics[width=0.5\textwidth]{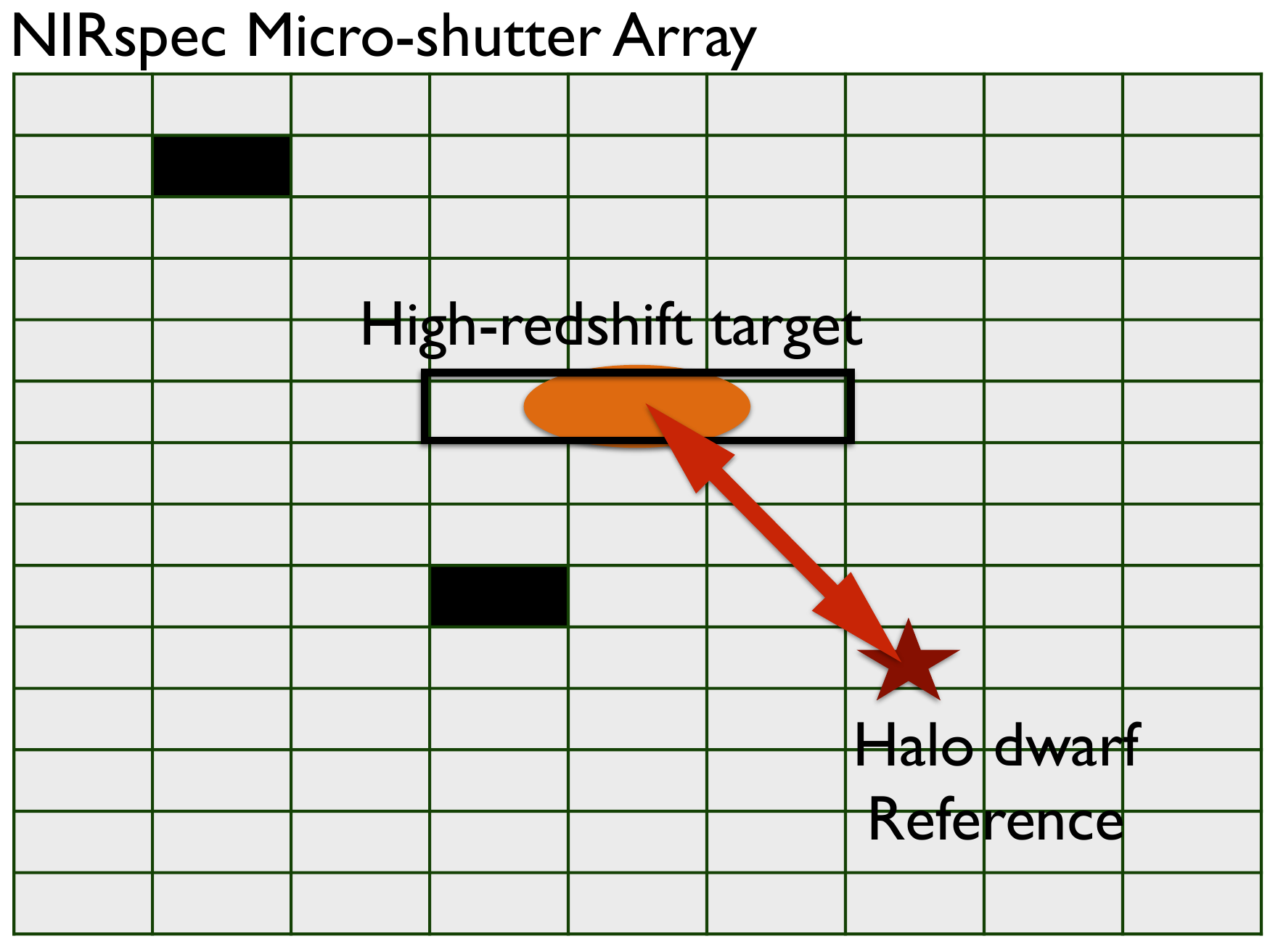}
	\caption{A cartoon of the NIRspec single-quadrant registration. In order to position the MSA slitlet on a target (e.g., a high-redshift galaxy) within the tolerance, it may be desirable to identify a nearby star in the pre-image within the same MSA quadrant (see Figure \ref{f:fov}).}
	\label{f:cartoon}
\end{center}
\end{figure}

\begin{figure}
\begin{center}
	\includegraphics[width=0.5\textwidth]{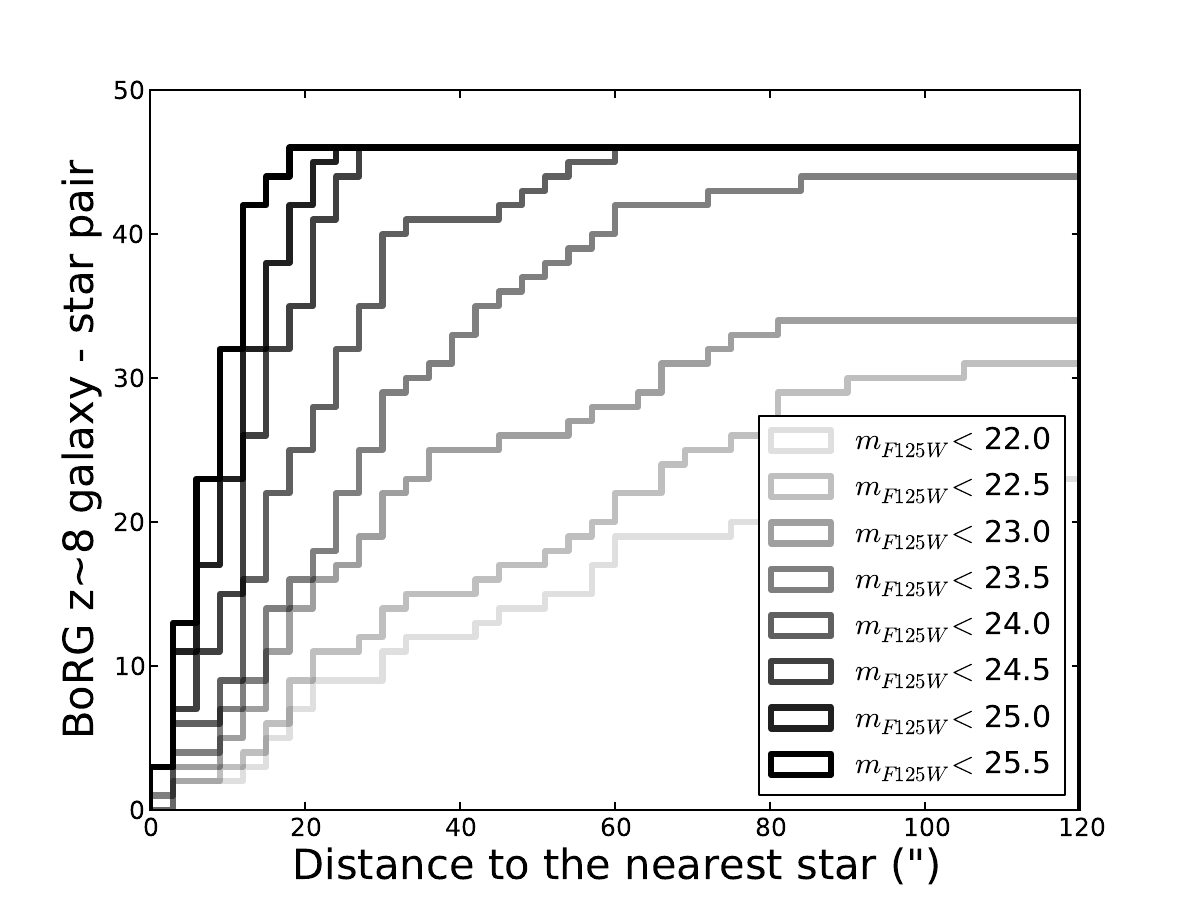}
\caption{The distribution of on-sky distance between a target $z\sim8$ galaxy and the nearest stars in the BoRG[z8] field color-coded by limiting magnitude (the limit to which stars are selected, typically a magnitude above the detection limit). }
\label{f:minmu}
\end{center}
\end{figure}

\subsection{Will we know the position of the scientific targets with respect to the reference stars with a high-enough accuracy? }
\label{s:neareststar}

Given a science target, how far away will the nearest reference star be? The BoRG[z8] fields are ideal for this experiment as their science target is $z\sim8$ galaxies, randomly distributed throughout the fields and we have already identified the stars. To accurately position the NIRspec MSA, we may not want or need to position the {\em absolute} astrometry but just the relative one. In this case it becomes important where the nearest star will likely be. To explore this, we calculate the on-sky separations between the nominal NIRspec targets, the $z\sim8$ candidate galaxies and the stars identified in all the BoRG[z8] fields.

Figure \ref{f:minmu} shows the distribution of on-sky distances to the nearest star for the $z\sim8$ galaxies. 
If the requirement is that a guide star is very close to the target, one is practically guaranteed to be within 20" if the limiting magnitude of the pre-image is $m_{F125W} < 25.5$ (93\% of all $z\sim8$ targets).
% In most fields, a star can be found between $\mu=5$" and $\mu=15$" ($m_{F125W} < 25.5$) and all $z\sim8$ galaxies have a guide star within $\mu=20$". We note that this distance is not limited by the WFC3 FOV or the NIRspec one.
However, if one requires the nearest guide star to be within the MSA quadrant assigned to the target, the limiting magnitude of the pre-spectroscopy image will need to be  $m_{F125W}  = 24-24.5$, placing the nearest reference star within $\sim$45". 

% Figure \ref{f:mlim} explores this further with the mean minimum distance between a $z\sim8$ target galaxy and a star as a function of the limiting magnitude for which stars are identified. We note that this limiting magnitude for stellar identification is typically about a magnitude above the limiting magnitude for object detection. 
%
%Figure \ref{f:allmu} shows the on-sky distances between a target $z\sim8$ galaxy and {\em all} the stars in a WFC3 field, combined over all the fields. 
%It shows that within 20", there are typically already some of stars with $m_{F125W} < 25.5$ available to register the image. 
%
%

\section{Concluding Remarks}
\label{s:concl}

We explored the number of stars available to astrometrically register the on-sky position of NIRspec observations using a pre-spectroscopic image. 
Our approach was to use WFC3 near-infrared images from the BoRG[z8] survey with known high-redshift targets and positive identifications of Galactic dwarfs to explore if there are enough such stars in a WFC3 or NIRspec field-of-view, if their number depends significantly on limiting depth or Galactic coordinates.

From the number of stars in the BoRG[z8] fields, their characteristics and relative position with respect to the $z\sim8$ galaxies, we can conclude the following:
\begin{itemize}
\item The number of Galactic stars suitable for astrometric reference in a NIRspec image is a strong function of limiting depth (Figures \ref{f:histNstar} and \ref{f:histNbrightstar}). 

\item The surface density of Galactic stars does not change as a function of Galactic coordinates (minimally $log_{10}(\#/arcmin^2) \sim1.5$ available, half a dex less for M-dwarfs) but the stellar number density shows strong field-to-field variations (Figure \ref{f:NstarGal}) but no dependence on Galactic coordinates above $|b| > 10^\circ$.

\item The surface density of Galactic stars in a typical single higher-latitude WFC3 near-infrared image is enough for successful NIRspec Target Acquisition (10\% MSA slit-width astrometry) provided the TA image is 24-24.5 AB mag ({\em F125W}) in depth (Figure \ref{f:FieldReg}). 
We note again here that this scenario is more appropriate for TA with the {\em F110W} filter and the {\em F140X} will perform much better as it is 6x wider than {\em F125W} and spans H-band, in which brown and red dwarfs are much brighter. Alternatively, one would need to know the reference stars types better in order to predict their relative luminosities in the {\em F140X} filter to be passed on to the centroiding algorithm (e.g. just use M-dwarfs identified from their colors for TA).

\item The NIRspec Target Acquisition can be based on a WFC3 mosaic if the mosaic and TA image has a depth 23 AB mag (F125W, Figure \ref{f:MosReg}).

\item If a requirement is that there is a single guide star close (angular separation $<20$") to a target, the pre-image will need a limiting depth of $m_{F125W} < 25.5$ (93\% of all $z\sim8$ sources have a nearby guide star). If the requirement is that one should be in the same MSA quadrant, a limiting depth of $m_{F125W} < 24.0$ is needed to guarantee one such guide star (Figure \ref{f:minmu}).
\end{itemize}

% We note that the consideration if the targets identified in a single WFC3 are indeed high-value enough is a relative one and should be left to the JWST time allocation committees. 
We note that the consideration if the targets identified in a single WFC3 pointing are indeed high-value enough is a relative one and should be left to the JWST time allocation committees.  This research note was solely to explore whether or not it is technically feasible or not.
It is feasible but will require a longer than baseline TA image integration (typically to a depth of $\sim18AB$) but well within the range of possible integrations.

\subsection{Possible Future Use: MIRI Pointing Registration}
\label{s:miri}

M-dwarf near- to mid-infrared color is stable at $\Delta m\sim0.6$ for the nearby sample presented by \citet{Kirkpatrick11} with WISE detections.
In effect, the F160W LF for M-dwarfs would be shifted up by $\sim0.6$ magnitude to get the expected apparent 18 $\mu$m LF in MIRI pre-images.
Thus, we can be certain to identify the M-dwarfs in each field. L- and T-type dwarfs are intrinsically brighter in the mid-infrared \citep{Kirkpatrick11}
but they are rarer in the BoRG[z8] fields given their similar Galactic scale-height \citep[][]{Ryan11, Holwerda14} and fainter absolute luminosities.
Therefore, the BoRG[z8] field counts of Galactic stars can, in principle, be used to estimate contamination and pointing for the MIRI instrument as well.

\section*{Acknowledgements}

We would like to thank the referee for his or her comments and suggestions which helped improve this manuscript. 
%The authors would like tot thank N. Reid and K. Cruz for their help with the local density of M-dwarfs and R. Benjamin for useful discussions on Galactic substructures.
%
The lead author thanks the European Space Agency for the Fellowship program and its support and T. B\"{o}ker, P. Ferruit, G. Giardino, M. Franx, and T. Beck for fruitful discussions on the topic of NIRspec target acquisition. 
We acknowledge the support of HST Archival grant number AR-12134, General Observer Grants GO-11700, GO-12572, and G0-12905
and the European Space Agency for support of this work.
%
% MT  was partially supported by the European Commission through the Marie Curie Career Integration Fellowship PCIG12-GA-2012-333749"
%
This work is based in part on observations taken by the CANDELS Multi-Cycle Treasury Program with the NASA/ESA HST, which is operated by the Association of Universities for Research in Astronomy, Inc., under NASA contract NAS5-26555.
This research has made use of the NASA/IPAC Extragalactic Database (NED) which is operated by the Jet Propulsion Laboratory, California Institute of Technology, under contract with the National Aeronautics and Space Administration. 
This research has made use of NASA's Astrophysics Data System.
This research made use of Astropy, a community-developed core Python package for Astronomy \citep{Astropy-Collaboration13a}. This research made use of matplotlib, a Python library for publication quality graphics \citep{Hunter07}. PyRAF is a product of the Space Telescope Science Institute, which is operated by AURA for NASA. This research made use of SciPy \citep{scipy}.

%\clearpage

% note wise color of 100 M-dwarfs; useful for JWST transformation: \citet{Kirkpatrick11}


\begin{thebibliography}{33}
\newcommand{\enquote}[1]{``#1''}
\providecommand{\natexlab}[1]{#1}
\providecommand{\url}[1]{\texttt{#1}}
\providecommand{\urlprefix}{URL: }
\expandafter\ifx\csname urlstyle\endcsname\relax
  \providecommand{\doi}[1]{doi:\discretionary{}{}{}#1}\else
  \providecommand{\doi}{doi:\discretionary{}{}{}\begingroup
  \urlstyle{rm}\Url}\fi

\bibitem[{{Astropy Collaboration} \emph{et~al.}(2013){Astropy Collaboration},
  {Robitaille}, {Tollerud}, {Greenfield}, {Droettboom}, {Bray}, {Aldcroft},
  {Davis}, {Ginsburg}, {Price-Whelan}, {Kerzendorf}, {Conley}, {Crighton},
  {Barbary}, {Muna}, {Ferguson}, {Grollier}, {Parikh}, {Nair}, {Unther},
  {Deil}, {Woillez}, {Conseil}, {Kramer}, {Turner}, {Singer}, {Fox}, {Weaver},
  {Zabalza}, {Edwards}, {Azalee Bostroem}, {Burke}, {Casey}, {Crawford},
  {Dencheva}, {Ely}, {Jenness}, {Labrie}, {Lim}, {Pierfederici}, {Pontzen},
  {Ptak}, {Refsdal}, {Servillat} \& {Streicher}}]{Astropy-Collaboration13a}
{Astropy Collaboration}, {Robitaille}, T.~P., {Tollerud}, E.~J., {Greenfield},
  P., {Droettboom}, M., {Bray}, E., {Aldcroft}, T., {Davis}, M., {Ginsburg},
  A., {Price-Whelan}, A.~M., {Kerzendorf}, W.~E., {Conley}, A., {Crighton}, N.,
  {Barbary}, K., {Muna}, D., {Ferguson}, H., {Grollier}, F., {Parikh}, M.~M.,
  {Nair}, P.~H., {Unther}, H.~M., {Deil}, C., {Woillez}, J., {Conseil}, S.,
  {Kramer}, R., {Turner}, J.~E.~H., {Singer}, L., {Fox}, R., {Weaver}, B.~A.,
  {Zabalza}, V., {Edwards}, Z.~I., {Azalee Bostroem}, K., {Burke}, D.~J.,
  {Casey}, A.~R., {Crawford}, S.~M., {Dencheva}, N., {Ely}, J., {Jenness}, T.,
  {Labrie}, K., {Lim}, P.~L., {Pierfederici}, F., {Pontzen}, A., {Ptak}, A.,
  {Refsdal}, B., {Servillat}, M. \& {Streicher}, O. [2013]  \emph{\aap}
  \textbf{558}, A33, \doi{10.1051/0004-6361/201322068}.

\bibitem[{{Beck}(2009)}]{Beck09}
{Beck}, T. [2009]  \enquote{Alternative strategies for nirspec target
  acquisition in fs and ifu modes,} Tech. rep., STSCI.

\bibitem[{{Beck} \emph{et~al.}(2016){Beck}, {Karakla}, {Keyes}, {Ubeda} \& {the
  STScI NIRSpec Team}}]{Beck15}
{Beck}, T., {Karakla}, D., {Keyes}, D., {Ubeda}, L. \& {the STScI NIRSpec Team}
  [2016]  \enquote{Integral field spectroscopy of embedded galactic
  protostars,}
  \urlprefix\url{http://www.cosmos.esa.int/documents/739790/758039/M11-TBeck.p%
df/183b9830-7855-4177-9af4-4fcdfca60e47}.

\bibitem[{{Bertin} \& {Arnouts}(1996)}]{SE}
{Bertin}, E. \& {Arnouts}, S. [1996]  \emph{\aaps} \textbf{117},  393, provided
  by the NASA Astrophysics Data System.

\bibitem[{{Bradley} \emph{et~al.}(2012){Bradley}, {Trenti}, {Oesch},
  {Stiavelli}, {Treu}, {Bouwens}, {Shull}, {Holwerda} \& {Pirzkal}}]{Bradley12}
{Bradley}, L.~D., {Trenti}, M., {Oesch}, P.~A., {Stiavelli}, M., {Treu}, T.,
  {Bouwens}, R.~J., {Shull}, J.~M., {Holwerda}, B.~W. \& {Pirzkal}, N. [2012]
  \emph{\apj} \textbf{760}, 108, \doi{10.1088/0004-637X/760/2/108}.

\bibitem[{{Calvi} \emph{et~al.}(2016){Calvi}, {Trenti}, {Stiavelli}, {Oesch},
  {Bradley}, {Schmidt}, {Coe}, {Brammer}, {Bernard}, {Bouwens}, {Carrasco},
  {Carollo}, {Holwerda}, {MacKenty}, {Mason}, {Shull} \& {Treu}}]{Calvi16}
{Calvi}, V., {Trenti}, M., {Stiavelli}, M., {Oesch}, P., {Bradley}, L.~D.,
  {Schmidt}, K.~B., {Coe}, D., {Brammer}, G., {Bernard}, S., {Bouwens}, R.~J.,
  {Carrasco}, D., {Carollo}, C.~M., {Holwerda}, B.~W., {MacKenty}, J.~W.,
  {Mason}, C.~A., {Shull}, J.~M. \& {Treu}, T. [2016]  \emph{\apj}
  \textbf{817}, 120, \doi{10.3847/0004-637X/817/2/120}.

\bibitem[{{Casertano} \emph{et~al.}(2000){Casertano}, {de Mello}, {Dickinson},
  {Ferguson}, {Fruchter}, {Gonzalez-Lopezlira}, {Heyer}, {Hook}, {Levay},
  {Lucas}, {Mack}, {Makidon}, {Mutchler}, {Smith}, {Stiavelli}, {Wiggs} \&
  {Williams}}]{Casertano00}
{Casertano}, S., {de Mello}, D., {Dickinson}, M., {Ferguson}, H.~C.,
  {Fruchter}, A.~S., {Gonzalez-Lopezlira}, R.~A., {Heyer}, I., {Hook}, R.~N.,
  {Levay}, Z., {Lucas}, R.~A., {Mack}, J., {Makidon}, R.~B., {Mutchler}, M.,
  {Smith}, T.~E., {Stiavelli}, M., {Wiggs}, M.~S. \& {Williams}, R.~E. [2000]
  \emph{\aj} \textbf{120},  2747, \doi{10.1086/316851}.

\bibitem[{{Dressel} \emph{et~al.}(2010){Dressel}, {Wong} \&
  {Pavlovsky}}]{Dressel10}
{Dressel}, L., {Wong}, M.~H. \& {Pavlovsky}, C. [2010]  \enquote{Wide field
  camera 3 instrument handbook, version 3.0,}  \emph{Wide Field Camera 3
  Instrument Handbook, Version 3.0}.

\bibitem[{{Giavalisco} \emph{et~al.}(2004){Giavalisco}, {Ferguson},
  {Koekemoer}, {Dickinson}, {Alexander}, {Bauer}, {Bergeron}, {Biagetti},
  {Brandt}, {Casertano}, {Cesarsky}, {Chatzichristou}, {Conselice},
  {Cristiani}, {Da Costa}, {Dahlen}, {de Mello}, {Eisenhardt}, {Erben}, {Fall},
  {Fassnacht}, {Fosbury}, {Fruchter}, {Gardner}, {Grogin}, {Hook},
  {Hornschemeier}, {Idzi}, {Jogee}, {Kretchmer}, {Laidler}, {Lee}, {Livio},
  {Lucas}, {Madau}, {Mobasher}, {Moustakas}, {Nonino}, {Padovani}, {Papovich},
  {Park}, {Ravindranath}, {Renzini}, {Richardson}, {Riess}, {Rosati},
  {Schirmer}, {Schreier}, {Somerville}, {Spinrad}, {Stern}, {Stiavelli},
  {Strolger}, {Urry}, {Vandame}, {Williams} \& {Wolf}}]{goods}
{Giavalisco}, M., {Ferguson}, H.~C., {Koekemoer}, A.~M., {Dickinson}, M.,
  {Alexander}, D.~M., {Bauer}, F.~E., {Bergeron}, J., {Biagetti}, C., {Brandt},
  W.~N., {Casertano}, S., {Cesarsky}, C., {Chatzichristou}, E., {Conselice},
  C., {Cristiani}, S., {Da Costa}, L., {Dahlen}, T., {de Mello}, D.,
  {Eisenhardt}, P., {Erben}, T., {Fall}, S.~M., {Fassnacht}, C., {Fosbury}, R.,
  {Fruchter}, A., {Gardner}, J.~P., {Grogin}, N., {Hook}, R.~N.,
  {Hornschemeier}, A.~E., {Idzi}, R., {Jogee}, S., {Kretchmer}, C., {Laidler},
  V., {Lee}, K.~S., {Livio}, M., {Lucas}, R., {Madau}, P., {Mobasher}, B.,
  {Moustakas}, L.~A., {Nonino}, M., {Padovani}, P., {Papovich}, C., {Park}, Y.,
  {Ravindranath}, S., {Renzini}, A., {Richardson}, M., {Riess}, A., {Rosati},
  P., {Schirmer}, M., {Schreier}, E., {Somerville}, R.~S., {Spinrad}, H.,
  {Stern}, D., {Stiavelli}, M., {Strolger}, L., {Urry}, C.~M., {Vandame}, B.,
  {Williams}, R. \& {Wolf}, C. [2004]  \emph{\apjl} \textbf{600},  L93,
  \doi{10.1086/379232}.

\bibitem[{{Grogin} \emph{et~al.}(2011){Grogin}, {Kocevski}, {Faber},
  {Ferguson}, {Koekemoer}, {Riess}, {Acquaviva}, {Alexander}, {Almaini},
  {Ashby}, {Barden}, {Bell}, {Bournaud}, {Brown}, {Caputi}, {Casertano},
  {Cassata}, {Castellano}, {Challis}, {Chary}, {Cheung}, {Cirasuolo},
  {Conselice}, {Roshan Cooray}, {Croton}, {Daddi}, {Dahlen}, {Dav{\'e}}, {de
  Mello}, {Dekel}, {Dickinson}, {Dolch}, {Donley}, {Dunlop}, {Dutton}, {Elbaz},
  {Fazio}, {Filippenko}, {Finkelstein}, {Fontana}, {Gardner}, {Garnavich},
  {Gawiser}, {Giavalisco}, {Grazian}, {Guo}, {Hathi}, {H{\"a}ussler},
  {Hopkins}, {Huang}, {Huang}, {Jha}, {Kartaltepe}, {Kirshner}, {Koo}, {Lai},
  {Lee}, {Li}, {Lotz}, {Lucas}, {Madau}, {McCarthy}, {McGrath}, {McIntosh},
  {McLure}, {Mobasher}, {Moustakas}, {Mozena}, {Nandra}, {Newman}, {Niemi},
  {Noeske}, {Papovich}, {Pentericci}, {Pope}, {Primack}, {Rajan},
  {Ravindranath}, {Reddy}, {Renzini}, {Rix}, {Robaina}, {Rodney}, {Rosario},
  {Rosati}, {Salimbeni}, {Scarlata}, {Siana}, {Simard}, {Smidt}, {Somerville},
  {Spinrad}, {Straughn}, {Strolger}, {Telford}, {Teplitz}, {Trump}, {van der
  Wel}, {Villforth}, {Wechsler}, {Weiner}, {Wiklind}, {Wild}, {Wilson},
  {Wuyts}, {Yan} \& {Yun}}]{Grogin11}
{Grogin}, N.~A., {Kocevski}, D.~D., {Faber}, S.~M., {Ferguson}, H.~C.,
  {Koekemoer}, A.~M., {Riess}, A.~G., {Acquaviva}, V., {Alexander}, D.~M.,
  {Almaini}, O., {Ashby}, M.~L.~N., {Barden}, M., {Bell}, E.~F., {Bournaud},
  F., {Brown}, T.~M., {Caputi}, K.~I., {Casertano}, S., {Cassata}, P.,
  {Castellano}, M., {Challis}, P., {Chary}, R.-R., {Cheung}, E., {Cirasuolo},
  M., {Conselice}, C.~J., {Roshan Cooray}, A., {Croton}, D.~J., {Daddi}, E.,
  {Dahlen}, T., {Dav{\'e}}, R., {de Mello}, D.~F., {Dekel}, A., {Dickinson},
  M., {Dolch}, T., {Donley}, J.~L., {Dunlop}, J.~S., {Dutton}, A.~A., {Elbaz},
  D., {Fazio}, G.~G., {Filippenko}, A.~V., {Finkelstein}, S.~L., {Fontana}, A.,
  {Gardner}, J.~P., {Garnavich}, P.~M., {Gawiser}, E., {Giavalisco}, M.,
  {Grazian}, A., {Guo}, Y., {Hathi}, N.~P., {H{\"a}ussler}, B., {Hopkins},
  P.~F., {Huang}, J.-S., {Huang}, K.-H., {Jha}, S.~W., {Kartaltepe}, J.~S.,
  {Kirshner}, R.~P., {Koo}, D.~C., {Lai}, K., {Lee}, K.-S., {Li}, W., {Lotz},
  J.~M., {Lucas}, R.~A., {Madau}, P., {McCarthy}, P.~J., {McGrath}, E.~J.,
  {McIntosh}, D.~H., {McLure}, R.~J., {Mobasher}, B., {Moustakas}, L.~A.,
  {Mozena}, M., {Nandra}, K., {Newman}, J.~A., {Niemi}, S.-M., {Noeske}, K.~G.,
  {Papovich}, C.~J., {Pentericci}, L., {Pope}, A., {Primack}, J.~R., {Rajan},
  A., {Ravindranath}, S., {Reddy}, N.~A., {Renzini}, A., {Rix}, H.-W.,
  {Robaina}, A.~R., {Rodney}, S.~A., {Rosario}, D.~J., {Rosati}, P.,
  {Salimbeni}, S., {Scarlata}, C., {Siana}, B., {Simard}, L., {Smidt}, J.,
  {Somerville}, R.~S., {Spinrad}, H., {Straughn}, A.~N., {Strolger}, L.-G.,
  {Telford}, O., {Teplitz}, H.~I., {Trump}, J.~R., {van der Wel}, A.,
  {Villforth}, C., {Wechsler}, R.~H., {Weiner}, B.~J., {Wiklind}, T., {Wild},
  V., {Wilson}, G., {Wuyts}, S., {Yan}, H.-J. \& {Yun}, M.~S. [2011]
  \emph{\apjs} \textbf{197}, 35, \doi{10.1088/0067-0049/197/2/35}.

\bibitem[{{Holwerda}(2005{\natexlab{a}})}]{seman}
{Holwerda}, B.~W. [2005{\natexlab{a}}]  \emph{astro-ph/0512139} .

\bibitem[{{Holwerda}(2005{\natexlab{b}})}]{mythesis}
{Holwerda}, B.~W. [2005{\natexlab{b}}]  \enquote{{The opacity of spiral galaxy
  disks},}  PhD thesis, Proefschrift, Rijksuniversiteit Groningen, 2005.

\bibitem[{{Holwerda} \emph{et~al.}(2005){Holwerda}, {Gonz\'alez}, {Allen} \&
  {van der Kruit}}]{Holwerda05a}
{Holwerda}, B.~W., {Gonz\'alez}, R.~A., {Allen}, R.~J. \& {van der Kruit},
  P.~C. [2005]  \emph{\aj} \textbf{129},  1381.

\bibitem[{{Holwerda} \emph{et~al.}(2014){Holwerda}, {Trenti}, {Clarkson},
  {Sahu}, {Bradley}, {Stiavelli}, {Pirzkal}, {De Marchi}, {Andersen}, {Bouwens}
  \& {Ryan}}]{Holwerda14}
{Holwerda}, B.~W., {Trenti}, M., {Clarkson}, W., {Sahu}, K., {Bradley}, L.,
  {Stiavelli}, M., {Pirzkal}, N., {De Marchi}, G., {Andersen}, M., {Bouwens},
  R. \& {Ryan}, R. [2014]  \emph{\apj} \textbf{788}, 77,
  \doi{10.1088/0004-637X/788/1/77}.

\bibitem[{Hunter(2007)}]{Hunter07}
Hunter, J.~D. [2007]  \emph{Computing In Science \& Engineering} \textbf{9},
  90.

\bibitem[{{Jones} \emph{et~al.}(2001){Jones}, {Oliphant}, {Peterson} \&
  Others}]{scipy}
{Jones}, E., {Oliphant}, T., {Peterson}, P. \& Others [2001]  \enquote{{SciPy}:
  Open source scientific tools for python,}
  \urlprefix\url{http://www.scipy.org/}.

\bibitem[{{Karakla} \emph{et~al.}(2015){Karakla}, {Beck}, {Gilbert} \& {the
  STScI NIRSpec Team}}]{Karakla15}
{Karakla}, D., {Beck}, T., {Gilbert}, G., K. amd~{Curtis} \& {the STScI NIRSpec
  Team} [2015]  \enquote{A demonstration of the nirspec micro-shutter array
  planning software: Observing high redshift galaxies in the hudf with jwst,}
  \urlprefix\url{http://www.cosmos.esa.int/documents/739790/758039/M04-DKarakl%
a.pdf/582c2806-1283-46ea-8533-3371c38e81cf}.

\bibitem[{Kirkpatrick \emph{et~al.}(2011)Kirkpatrick, Cushing, Gelino,
  Griffith, Skrutskie, Marsh, Wright, Mainzer, Eisenhardt, McLean, Thompson,
  Bauer, Benford, Bridge, Lake, Petty, Stanford, Tsai, Bailey, Beichman, Bloom,
  Bochanski, Burgasser, Capak, Cruz, Hinz, Kartaltepe, Knox, Manohar, Masters,
  Morales-Calder{\'o}n, Prato, Rodigas, Salvato, Schurr, Scoville, Simcoe,
  Stapelfeldt, Stern, Stock \& Vacca}]{Kirkpatrick11}
Kirkpatrick, J.~D., Cushing, M.~C., Gelino, C.~R., Griffith, R.~L., Skrutskie,
  M.~F., Marsh, K.~A., Wright, E.~L., Mainzer, A., Eisenhardt, P.~R., McLean,
  I.~S., Thompson, M.~A., Bauer, J.~M., Benford, D.~J., Bridge, C.~R., Lake,
  S.~E., Petty, S.~M., Stanford, S.~A., Tsai, C.-W., Bailey, V., Beichman,
  C.~A., Bloom, J.~S., Bochanski, J.~J., Burgasser, A.~J., Capak, P.~L., Cruz,
  K.~L., Hinz, P.~M., Kartaltepe, J.~S., Knox, R.~P., Manohar, S., Masters, D.,
  Morales-Calder{\'o}n, M., Prato, L.~A., Rodigas, T.~J., Salvato, M., Schurr,
  S.~D., Scoville, N.~Z., Simcoe, R.~A., Stapelfeldt, K.~R., Stern, D., Stock,
  N.~D. \& Vacca, W.~D. [2011]  \emph{The Astrophysical Journal Supplement
  Series} \textbf{197},  19,
  \urlprefix\url{http://stacks.iop.org/0067-0049/197/i=2/a=19}.

\bibitem[{{Koekemoer} \emph{et~al.}(2011){Koekemoer}, {Faber}, {Ferguson},
  {Grogin}, {Kocevski}, {Koo}, {Lai}, {Lotz}, {Lucas}, {McGrath}, {Ogaz},
  {Rajan}, {Riess}, {Rodney}, {Strolger}, {Casertano}, {Castellano}, {Dahlen},
  {Dickinson}, {Dolch}, {Fontana}, {Giavalisco}, {Grazian}, {Guo}, {Hathi},
  {Huang}, {van der Wel}, {Yan}, {Acquaviva}, {Alexander}, {Almaini}, {Ashby},
  {Barden}, {Bell}, {Bournaud}, {Brown}, {Caputi}, {Cassata}, {Challis},
  {Chary}, {Cheung}, {Cirasuolo}, {Conselice}, {Roshan Cooray}, {Croton},
  {Daddi}, {Dav{\'e}}, {de Mello}, {de Ravel}, {Dekel}, {Donley}, {Dunlop},
  {Dutton}, {Elbaz}, {Fazio}, {Filippenko}, {Finkelstein}, {Frazer}, {Gardner},
  {Garnavich}, {Gawiser}, {Gruetzbauch}, {Hartley}, {H{\"a}ussler},
  {Herrington}, {Hopkins}, {Huang}, {Jha}, {Johnson}, {Kartaltepe},
  {Khostovan}, {Kirshner}, {Lani}, {Lee}, {Li}, {Madau}, {McCarthy},
  {McIntosh}, {McLure}, {McPartland}, {Mobasher}, {Moreira}, {Mortlock},
  {Moustakas}, {Mozena}, {Nandra}, {Newman}, {Nielsen}, {Niemi}, {Noeske},
  {Papovich}, {Pentericci}, {Pope}, {Primack}, {Ravindranath}, {Reddy},
  {Renzini}, {Rix}, {Robaina}, {Rosario}, {Rosati}, {Salimbeni}, {Scarlata},
  {Siana}, {Simard}, {Smidt}, {Snyder}, {Somerville}, {Spinrad}, {Straughn},
  {Telford}, {Teplitz}, {Trump}, {Vargas}, {Villforth}, {Wagner}, {Wandro},
  {Wechsler}, {Weiner}, {Wiklind}, {Wild}, {Wilson}, {Wuyts} \&
  {Yun}}]{koekemoer11}
{Koekemoer}, A.~M., {Faber}, S.~M., {Ferguson}, H.~C., {Grogin}, N.~A.,
  {Kocevski}, D.~D., {Koo}, D.~C., {Lai}, K., {Lotz}, J.~M., {Lucas}, R.~A.,
  {McGrath}, E.~J., {Ogaz}, S., {Rajan}, A., {Riess}, A.~G., {Rodney}, S.~A.,
  {Strolger}, L., {Casertano}, S., {Castellano}, M., {Dahlen}, T., {Dickinson},
  M., {Dolch}, T., {Fontana}, A., {Giavalisco}, M., {Grazian}, A., {Guo}, Y.,
  {Hathi}, N.~P., {Huang}, K.-H., {van der Wel}, A., {Yan}, H.-J., {Acquaviva},
  V., {Alexander}, D.~M., {Almaini}, O., {Ashby}, M.~L.~N., {Barden}, M.,
  {Bell}, E.~F., {Bournaud}, F., {Brown}, T.~M., {Caputi}, K.~I., {Cassata},
  P., {Challis}, P.~J., {Chary}, R.-R., {Cheung}, E., {Cirasuolo}, M.,
  {Conselice}, C.~J., {Roshan Cooray}, A., {Croton}, D.~J., {Daddi}, E.,
  {Dav{\'e}}, R., {de Mello}, D.~F., {de Ravel}, L., {Dekel}, A., {Donley},
  J.~L., {Dunlop}, J.~S., {Dutton}, A.~A., {Elbaz}, D., {Fazio}, G.~G.,
  {Filippenko}, A.~V., {Finkelstein}, S.~L., {Frazer}, C., {Gardner}, J.~P.,
  {Garnavich}, P.~M., {Gawiser}, E., {Gruetzbauch}, R., {Hartley}, W.~G.,
  {H{\"a}ussler}, B., {Herrington}, J., {Hopkins}, P.~F., {Huang}, J.-S.,
  {Jha}, S.~W., {Johnson}, A., {Kartaltepe}, J.~S., {Khostovan}, A.~A.,
  {Kirshner}, R.~P., {Lani}, C., {Lee}, K.-S., {Li}, W., {Madau}, P.,
  {McCarthy}, P.~J., {McIntosh}, D.~H., {McLure}, R.~J., {McPartland}, C.,
  {Mobasher}, B., {Moreira}, H., {Mortlock}, A., {Moustakas}, L.~A., {Mozena},
  M., {Nandra}, K., {Newman}, J.~A., {Nielsen}, J.~L., {Niemi}, S., {Noeske},
  K.~G., {Papovich}, C.~J., {Pentericci}, L., {Pope}, A., {Primack}, J.~R.,
  {Ravindranath}, S., {Reddy}, N.~A., {Renzini}, A., {Rix}, H.-W., {Robaina},
  A.~R., {Rosario}, D.~J., {Rosati}, P., {Salimbeni}, S., {Scarlata}, C.,
  {Siana}, B., {Simard}, L., {Smidt}, J., {Snyder}, D., {Somerville}, R.~S.,
  {Spinrad}, H., {Straughn}, A.~N., {Telford}, O., {Teplitz}, H.~I., {Trump},
  J.~R., {Vargas}, C., {Villforth}, C., {Wagner}, C.~R., {Wandro}, P.,
  {Wechsler}, R.~H., {Weiner}, B.~J., {Wiklind}, T., {Wild}, V., {Wilson}, G.,
  {Wuyts}, S. \& {Yun}, M.~S. [2011]  \emph{\apjs} \textbf{197}, 36,
  \doi{10.1088/0067-0049/197/2/36}.

\bibitem[{{Pirzkal} \emph{et~al.}(2009){Pirzkal}, {Burgasser}, {Malhotra},
  {Holwerda}, {Sahu}, {Rhoads}, {Xu}, {Bochanski}, {Walsh}, {Windhorst},
  {Hathi} \& {Cohen}}]{Pirzkal09}
{Pirzkal}, N., {Burgasser}, A.~J., {Malhotra}, S., {Holwerda}, B.~W., {Sahu},
  K.~C., {Rhoads}, J.~E., {Xu}, C., {Bochanski}, J.~J., {Walsh}, J.~R.,
  {Windhorst}, R.~A., {Hathi}, N.~P. \& {Cohen}, S.~H. [2009]  \emph{\apj}
  \textbf{695},  1591, \doi{10.1088/0004-637X/695/2/1591}.

\bibitem[{{Radburn-Smith} \emph{et~al.}(2011){Radburn-Smith}, {de Jong},
  {Seth}, {Bailin}, {Bell}, {Brown}, {Bullock}, {Courteau}, {Dalcanton},
  {Ferguson}, {Goudfrooij}, {Holfeltz}, {Holwerda}, {Purcell}, {Sick},
  {Streich}, {Vlajic} \& {Zucker}}]{GHOSTS}
{Radburn-Smith}, D.~J., {de Jong}, R.~S., {Seth}, A.~C., {Bailin}, J., {Bell},
  E.~F., {Brown}, T.~M., {Bullock}, J.~S., {Courteau}, S., {Dalcanton}, J.~J.,
  {Ferguson}, H.~C., {Goudfrooij}, P., {Holfeltz}, S., {Holwerda}, B.~W.,
  {Purcell}, C., {Sick}, J., {Streich}, D., {Vlajic}, M. \& {Zucker}, D.~B.
  [2011]  \emph{\apjs} \textbf{195},  18, \doi{10.1088/0067-0049/195/2/18}.

\bibitem[{{Regan}(2005)}]{Regan05}
{Regan}, M. [2005]  \enquote{An alternative observing strategy for nirspec and
  its effect on nirspec target acquisition,} Tech. Rep. STScI-000674, STSCI,
  \urlprefix\url{http://www.stsci.edu/jwst/instruments/nirspec/docarchive/JWST%
-STScI-000674.pdf
  http://www.stsci.edu/jwst/instruments/nirspec/docarchive/JWST-STScI-000674.p%
df}.

\bibitem[{{Robin} \emph{et~al.}(2003){Robin}, {Reyl{\'e}}, {Derri{\`e}re} \&
  {Picaud}}]{Robin03}
{Robin}, A.~C., {Reyl{\'e}}, C., {Derri{\`e}re}, S. \& {Picaud}, S. [2003]
  \emph{\aap} \textbf{409},  523, \doi{10.1051/0004-6361:20031117}.

\bibitem[{{Ryan} \emph{et~al.}(2011){Ryan}, {Thorman}, {Yan}, {Fan}, {Yan},
  {Mechtley}, {Hathi}, {Cohen}, {Windhorst}, {McCarthy} \& {Wittman}}]{Ryan11}
{Ryan}, R.~E., {Thorman}, P.~A., {Yan}, H., {Fan}, X., {Yan}, L., {Mechtley},
  M.~R., {Hathi}, N.~P., {Cohen}, S.~H., {Windhorst}, R.~A., {McCarthy}, P.~J.
  \& {Wittman}, D.~M. [2011]  \emph{\apj} \textbf{739}, 83,
  \doi{10.1088/0004-637X/739/2/83}.

\bibitem[{{Schmidt} \emph{et~al.}(2014){Schmidt}, {Treu}, {Trenti}, {Bradley},
  {Kelly}, {Oesch}, {Holwerda}, {Shull} \& {Stiavelli}}]{Schmidt14}
{Schmidt}, K.~B., {Treu}, T., {Trenti}, M., {Bradley}, L.~D., {Kelly}, B.~C.,
  {Oesch}, P.~A., {Holwerda}, B.~W., {Shull}, J.~M. \& {Stiavelli}, M. [2014]
  \emph{\apj} \textbf{786}, 57, \doi{10.1088/0004-637X/786/1/57}.

\bibitem[{{Steidel} \emph{et~al.}(1996){Steidel}, {Giavalisco}, {Dickinson} \&
  {Adelberger}}]{Steidel96}
{Steidel}, C.~C., {Giavalisco}, M., {Dickinson}, M. \& {Adelberger}, K.~L.
  [1996]  \emph{\aj} \textbf{112},  352, \doi{10.1086/118019}.

\bibitem[{{Trenti} \emph{et~al.}(2011){Trenti}, {Bradley}, {Stiavelli},
  {Oesch}, {Treu}, {Bouwens}, {Shull}, {MacKenty}, {Carollo} \&
  {Illingworth}}]{Trenti11}
{Trenti}, M., {Bradley}, L.~D., {Stiavelli}, M., {Oesch}, P., {Treu}, T.,
  {Bouwens}, R.~J., {Shull}, J.~M., {MacKenty}, J.~W., {Carollo}, C.~M. \&
  {Illingworth}, G.~D. [2011]  \emph{\apjl} \textbf{727}, L39,
  \doi{10.1088/2041-8205/727/2/L39}.

\bibitem[{{Trenti} \& {Stiavelli}(2008)}]{Trenti08}
{Trenti}, M. \& {Stiavelli}, M. [2008]  \emph{\apj} \textbf{676},  767,
  \doi{10.1086/528674}.

\bibitem[{{Ubeda} \emph{et~al.}(2015){Ubeda}, {Beck} \& the STScI
  NIRSpec~Team}]{Ubeda15}
{Ubeda}, L., {Beck}, T. \& the STScI NIRSpec~Team [2015]  \enquote{Preparing
  nirspec observations: Nirspec pre-imaging using archival hst and simulated
  jwst/nircam data,}
  \urlprefix\url{http://www.cosmos.esa.int/documents/739790/758039/M12-LUbeda.%
pdf/a3d80490-5405-4a08-be71-f1b7ae00a337}.

\bibitem[{{van Dokkum}(2001)}]{van-Dokkum01}
{van Dokkum}, P.~G. [2001]  \emph{\pasp} \textbf{113},  1420,
  \doi{10.1086/323894}.

\bibitem[{{van Vledder} \emph{et~al.}(2016){van Vledder}, {van der Vlugt},
  {Holwerda}, {Kenworthy}, {Bouwens} \& {Trenti}}]{van-Vledder16}
{van Vledder}, I., {van der Vlugt}, D., {Holwerda}, B.~W., {Kenworthy}, M.~A.,
  {Bouwens}, R.~J. \& {Trenti}, M. [2016]  \emph{\mnras}
  \doi{10.1093/mnras/stw258}.

\bibitem[{{Windhorst} \emph{et~al.}(2011){Windhorst}, {Cohen}, {Hathi},
  {McCarthy}, {Ryan}, {Yan}, {Baldry}, {Driver}, {Frogel}, {Hill}, {Kelvin},
  {Koekemoer}, {Mechtley}, {O'Connell}, {Robotham}, {Rutkowski}, {Seibert},
  {Straughn}, {Tuffs}, {Balick}, {Bond}, {Bushouse}, {Calzetti}, {Crockett},
  {Disney}, {Dopita}, {Hall}, {Holtzman}, {Kaviraj}, {Kimble}, {MacKenty},
  {Mutchler}, {Paresce}, {Saha}, {Silk}, {Trauger}, {Walker}, {Whitmore} \&
  {Young}}]{Windhorst11}
{Windhorst}, R.~A., {Cohen}, S.~H., {Hathi}, N.~P., {McCarthy}, P.~J., {Ryan},
  R.~E., {Yan}, H., {Baldry}, I.~K., {Driver}, S.~P., {Frogel}, J.~A., {Hill},
  D.~T., {Kelvin}, L.~S., {Koekemoer}, A.~M., {Mechtley}, M., {O'Connell},
  R.~W., {Robotham}, A.~S.~G., {Rutkowski}, M.~J., {Seibert}, M., {Straughn},
  A.~N., {Tuffs}, R.~J., {Balick}, B., {Bond}, H.~E., {Bushouse}, H.,
  {Calzetti}, D., {Crockett}, M., {Disney}, M.~J., {Dopita}, M.~A., {Hall},
  D.~N.~B., {Holtzman}, J.~A., {Kaviraj}, S., {Kimble}, R.~A., {MacKenty},
  J.~W., {Mutchler}, M., {Paresce}, F., {Saha}, A., {Silk}, J.~I., {Trauger},
  J.~T., {Walker}, A.~R., {Whitmore}, B.~C. \& {Young}, E.~T. [2011]
  \emph{\apjs} \textbf{193},  27, \doi{10.1088/0067-0049/193/2/27}.

\bibitem[{{Yan} \emph{et~al.}(2010){Yan}, {Yan}, {Zamojski}, {Windhorst},
  {McCarthy}, {Fan}, {R{\"o}ttgering}, {Koekemoer}, {Robertson}, {Dav{\'e}} \&
  {Cai}}]{Yan10b}
{Yan}, H., {Yan}, L., {Zamojski}, M.~A., {Windhorst}, R.~A., {McCarthy}, P.~J.,
  {Fan}, X., {R{\"o}ttgering}, H.~J.~A., {Koekemoer}, A.~M., {Robertson},
  B.~E., {Dav{\'e}}, R. \& {Cai}, Z. [2010]  \emph{ArXiv e-prints} .

\end{thebibliography}
\end{document}